\documentclass[amsmath,amssymb,nofootinbib,showpacs,twocolumn,nofootinbib]{revtex4-1}
\usepackage{color}
\definecolor{slateblue}{rgb}{0.2,0.22,0.6}
\usepackage{graphicx,amssymb,amsmath,times}
\usepackage{epsfig,epstopdf}
\usepackage[breaklinks=true,colorlinks=true,linkcolor=slateblue,citecolor=slateblue,urlcolor=slateblue,pdfauthor={Tomassetti},pdftitle={BCUncertainties}]{hyperref}
\usepackage{amsmath}
\usepackage{wasysym}

\def\Journal#1#2#3#4{{#4}, {#1}, {#2}, #3} 
\newcommand{\etal}{et al.}

\newcommand{\AMS}{\textsf{AMS}} 
\newcommand{\ApJ}{ApJ}
\newcommand{\AeA}{A\&A}
\newcommand{\ASR}{ASR}
\newcommand{\ApP}{Astropart. Phys.}
\newcommand{\PRD}{PRD}
\newcommand{\PRC}{PRC}
\newcommand{\PRL}{PRL}
\newcommand{\PLB}{PLB}

\newcommand{\myfont}{\textrm}
\newcommand{\p}{\ensuremath{p}}
\renewcommand{\d}{\myfont{d}}
\newcommand{\Hyd}{\myfont{H}}
\newcommand{\He}{\myfont{He}}
\newcommand{\Li}{\myfont{Li}}
\newcommand{\Be}{\myfont{Be}}
\newcommand{\B}{\myfont{B}}
\newcommand{\C}{\myfont{C}}
\newcommand{\N}{\myfont{N}}
\newcommand{\Oxy}{\myfont{O}}
\newcommand{\F}{\myfont{F}}
\newcommand{\Ne}{\myfont{Ne}}

\newcommand{\Si}{\myfont{Si}}

\newcommand{\K}{\myfont{K}}

\newcommand{\Fe}{\myfont{Fe}}

\newcommand{\BC}{\myfont{B}/\myfont{C}}

\newcommand{\GALPROP}{\textsf{GALPROP}}
\newcommand{\USINE}{\textsf{Usine}}
\newcommand{\SOLARPROP}{\textsf{SolarProp}}
\newcommand{\WNEW}{{\textsf{WNEW}}}
\newcommand{\YIELDX}{\textsf{YIELDX}}

\newcommand{\eplus}{\ensuremath{e^{\pm}}}
\newcommand{\pbar}{\ensuremath{\bar{p}}}
\newcommand{\dbar}{\ensuremath{\bar{d}}}

\newcommand{\R}{\mathcal{R}}
\newcommand{\ie}{\textit{i.e.}}
\newcommand{\eg}{\textit{e.g.}}
\newcommand{\Q}{\mathcal{Q}}
\newcommand{\XS}{{cross-section}}
\newcommand{\XSs}{{cross-sections}}
\newcommand{\PF}{\textsf{P}$\rightarrow$\textsf{F}}

\begin{document}

\title{Solar and nuclear physics uncertainties in cosmic-ray propagation}  

\author{Nicola Tomassetti}
\address{\it Universit{\`a} degli Studi di Perugia \& INFN-Perugia, I-06100 Perugia, Italy. \\E-mail: nicola.tomassetti@cern.ch}

\begin{abstract} 
%
Recent data released by the \AMS{} experiment on the primary spectra and secondary-to-primary ratios in cosmic rays (CRs)
can pose tight constraints to astrophysical models of CR acceleration and transport in the Galaxy,
thereby providing a robust baseline of the astrophysical background for dark matter search via antimatter.
However, models of CR propagation are affected by other important sources of uncertainties, notably from
solar modulation and nuclear fragmentation, that cannot be improved with the sole use of the \AMS{} data. 
The present work is aimed at assessing these uncertainties and their relevance in the interpretation of the new \AMS{} data on the boron-to-carbon (\BC) ratio.
Uncertainties from solar modulation are estimated using improved models of CR transport in the heliosphere
constrained against various types of measurements: monthly resolved CR data collected by balloon-born or space missions,
interstellar flux data from the Voyager-1 spacecraft, and counting rates from ground-based neutron monitor detectors.
Uncertainties from nuclear fragmentation are estimated using semiempirical cross-section formulae constrained by
measurements on isotopically resolved and charge-changing reactions.
We found that a proper data-driven treatment of solar modulation can guarantee the desired level of precision,
in comparison with the improved accuracy of the recent data on the \BC{} ratio.
On the other hand, nuclear uncertainties represent a serious limiting factor over a wide energy range.
We therefore stress the need for establishing a dedicated program of cross-section measurements at the $\mathcal{O}$(100\,GeV) energy scale.
\end{abstract}
\pacs{
  98.70.Sa, 
  96.50.S,  
  96.50.sh, 
  13.85.Tp, 
  95.35.+d} 
\maketitle

\section{Introduction}    
\label{Sec::Introduction} 
%
An important challenge in astroparticle physics is the determination of the injection and transport parameters 
of Galactic cosmic rays (CRs), that establish the relation between the measured fluxes near Earth and the properties of their sources.
Astrophysical models of CR propagation account for particle acceleration mechanisms in Galactic sources, 
diffusive transport processes in the turbulent magnetic fields, and production of secondary particles from CR collisions with the gas.
The vast majority of Galactic particles, such as protons, \He, and \C-\N-\Oxy{} nuclei are of \emph{primary} origin, 
originating by diffusive-shock acceleration in supernova remnants (SNRs). 
Rarer particles such as $^{2}$\Hyd, $^{3}$\He{} or \Li-\Be-\B{} elements are called s\emph{secondary}, as they are 
predominantly generated by interactions of primary CRs in the interstellar medium (ISM).
Along with the spectra of primary CRs, the use of secondary-to-primary nuclear ratios --and most notably the boron-to-carbon (\BC) ratio--
is of paramount importance for constraining the key parameters that regulate acceleration and transport properties of these particles \citep{Grenier2015}.
Understanding CR propagation is essential to predict the secondary production of antimatter particles \eplus, \pbar, or \dbar,
as it constitutes the \emph{astrophysical background} for the search of new-physics signals from the annihilation of dark matter particles.
Along with the background, the determination of the CR transport parameters is also important for modeling the signal of dark-matter induced antiparticles.
The \BC{} ratio in CRs has been precisely measured in the 0.5-1000\,GeV/n energy range by the Alpha Magnetic Spectrometer  (\AMS) experiment
in the \textit{International Space Station} \citep{Aguilar2016BC}.
Recent results from \AMS{} on the fluxes of protons and helium \citep{Aguilar2015Proton}, antiparticles \citep{Aguilar2016PbarP},
and preliminary measurements on high-energy \Li-\Be-\B{} spectra have generated novel ideas \citep{Serpico2015,TomassettiDonato2015,AmatoBlasi2017,Blasi2017}
and advanced analysis efforts \citep{Feng2016} aimed at the determination of subtle effects in CR propagation.

With these new standards of experimental accuracy, however, we consider critical to make an assessment of
those unavoidable sources of model uncertainties that cannot be directly constrained by the \AMS{} data:
solar-physics uncertainties, from the modulation effect of CRs in the heliosphere,
and nuclear-physics uncertainties, from fragmentation processes of light elements in the ISM.

The goal of this paper is to assess these uncertainties and their relevance in the interpretation of the \BC{} data.
Modeling solar modulation is of crucial importance to study the CR transport processes that 
reshape the CR spectra and secondary-to-primary ratios in the $\sim$\,0.1-20\,GeV/n energy region.
This problem is often underestimated in CR astrophysics, with the frequent use 
of ultrasimplified approaches based on poorly justified physics assumptions.
In this respect, a substantial advance can be done thanks to the availability of numerical solvers
for CR transport in the heliosphere and to the recent release of very precious sets of time-resolved data.
The second issue deals with nuclear physics inputs in CR propagation models.
The determination of the transport parameters relies on the calculation of the secondary 
production rate of \Be{} and \B{} nuclei, for which several energy-dependent cross-section estimates are required.
Propagation models make use of semiempirical \XS{} formulae that are based on accelerator data at the sub-GeV/n energy scale,
and then extrapolated at the relevant energies.
The accuracy of the inferred propagation parameters is clearly linked to the quality of
the available measurements on nuclear fragmentation.
Both sources of uncertainties are expected to affect the interpretation of secondary CR data, 
as they blur the connection between experimental observations and CR propagation effects,
therefore hiding valuable pieces of information that are encoded in the data.
A proper assessment of systematic uncertainties in the models is therefore essential, 
in CR physics analysis, for a full exploitation of the \AMS{} data.

\section{Observations}     
\label{Sec::Observations}  
%
In this section, we outline the various types of measurements that are used in this work. 
A reference model of CR propagation, which we use to provide local interstellar (IS) particle fluxes and nuclear ratios,
is constructed using recent primary CR data on proton and helium from \AMS{} \citep{Aguilar2015Proton} in the GeV/n - TeV/n energy region,
new measurements from CREAM-III \citep{Yoon2017} at multi-TeV/n energies, and direct IS flux data from Voyager-1 \citep{Cummings2016} in the sub-GeV/n energy window.
To model the \BC{} ratio, we make use of data released recently by \AMS{} and Voyager-1 \citep{Aguilar2016BC,Cummings2016}.
The \AMS{} measurements provide precise \BC{} ratio between 0.5 and 1000\,GeV/n of kinetic energy under a period of medium-high 
solar activity (from May 2011 to May 2016) corresponding to 2.3$\cdot$10$^{6}$ and 8.3$\cdot$10$^{6}$ boron and carbon events, respectively.
The Voyager-1 data are available only at 0.1\,GeV/n energies. Nonetheless, these data are very
precious because they constrain directly the low energy \BC{} ratio outside the heliosphere.

In order to model the solar modulation effect of CRs in the heliosphere, we made use of a large variety 
of data collected at Earth, in high-altitude balloons, or in space missions.
In particular, we use the data provided from the worldwide network of neutron monitor (NM) detectors.
We have retrieved monthly averaged measurements from NM stations in Newark, Oulu, Apatity, and Jungfraujoch \citep{Steigies2015}.
These data consist in energy-integrated counting rates, corrected for detector efficiency and atmospheric pressure, between January 2005 and January 2017.
NM rates are used in combination with direct measurements of the CR proton flux performed by space missions or high-altitude balloons projects.
We use monthly resolved proton data from the PAMELA experiment \citep{Adriani2013} collected under solar minimum conditions between July 2006 and January 2010, 
yearly-resolved data recently released by the EPHIN/SOHO satellite \citep{Kuhl2016}, collected over the period 2000-2016,
and proton flux measurements from the BESS Polar-I (Polar-II) mission between 13 and 21 December 2004 (between 23 December 2007 and 16 January 2008) \citep{Abe2012}.

Finally, we have made use of a large compilation of accelerator data on nuclear fragmentation reactions.
Most of the data consist in energy-dependent cross-sections for the production of $^{10}$\B, $^{11}$\B, $^{7}$\Be{}, $^{9}$\Be, and $^{10}$\Be{} 
isotopes from the collisions of \B-\C-\N-\Oxy{} elements off hydrogen.
These reactions have been measured between $\sim$\,50 to 5000\,MeV/n by a number of experimental projects
\citet{ReadViola1984,Webber1998,Webber1990c,Olson1983,Fontes1977,Korejwo2000,Korejwo2001,Radin1979,Ramaty1997,Webber1998prc,Raisbeck1971} (symbols are used later).
Along with reactions on isotopically separated projectiles and fragments, we also use measurements on charge-changing
reactions, from \citet{Webber2003}, which are available from $\sim$\,0.1\,GeV/n to nearly $100$\,GeV/n energies.
Finally, cross-sections involving heavier nuclei, such as $^{10}$\Ne, $^{28}$\Si, or $^{56}$\Fe, are
extracted from \GALPROP{} code of CR propagation \citep{StrongMoskalenko1998}.

\section{Reference model of CR propagation}  
\label{Sec::CRPropagation}                   
%
In this section, we define a reference model of CR diffusive propagation into an homogeneous cylindrical-shaped Galactic halo.
Models of CR propagation in the Galaxy employ fully analytical \citep{ThoudamHorandel2014,Tomassetti2012Hardening}. 
semianalytical \citep{Jones2001,Maurin2001}, or fully numerical calculation frameworks \citep{StrongMoskalenko1998,Evoli2008,Kissman2014}. 
We utilize semianalytical calculations implemented under \USINE, a global toolkit that allows to solve the CR propagation 
equation for given nuclear and astrophysical inputs such as ISM gas distribution, source distribution and spectra, 
fragmentation cross-sections, boundary conditions \citep{Maurin2015ICRC}.
We implement a Kraichnan-like diffusion model with minimal\--reacceleration as a benchmark.
The propagation equation for a  $j$-type CR particle is written as: 
\begin{align}\label{Eq::DiffusionTransport}
  \frac{\partial \mathcal{N}_{j}}{\partial t} = & Q^{\rm tot}_{j}+\vec{\nabla}\cdot\left(K\vec{\nabla}\mathcal{N}_j\right) - {\mathcal{N}_j}{\Gamma^{\rm tot}_{j}} \nonumber \\ 
+& \frac{\partial}{\partial \R} \R^{2} K_{RR} \frac{\partial}{\partial \R} \R^{2} \mathcal{N}_j - \frac{\partial}{\partial \R}\left(\dot{\R}_{j} \mathcal{N}_{j}\right) 
\end{align}
where $\mathcal{N}_{j}=dN_{j}/dVd\R$ is the CR density of the species $j$ per unit of rigidity $\R$. 
The source term, $Q^{\rm tot}_{j}=Q^{\rm pri}_{j} + Q^{\rm sec}_{j}$, includes the primary injection spectra, 
from SNRs, and the terms describing secondary CR production in the ISM or decays. 
The rigidity dependence of the primary term is traditionally modeled as a power law, $Q^{\rm pri}_{j} \propto \R^{-\nu}$,
with a universal injection index $\nu$ for all primary elements.
This picture, however, has recently been challenged by the observation of puzzling features in the spectrum of primary CRs \citep{Serpico2015,AmatoBlasi2017}:
a difference in spectral index between protons and $Z>1$ nuclei, with  \p/\He\,$\propto$\,$\R^{-0.08}$ at $\R\sim\,40$-$1800$\,GV,
and a common spectral hardening of all fluxes at $\R\gtrsim\R_{\rm B}\cong$\,375\,GV \citep{Aguilar2015Proton}.
Hence, we adopt a primary source term of the type:
%
\begin{table}[!t]
\begin{center}
\small
\begin{tabular}{lll}
\tableline
\tableline
Parameter & Name &  Value \\
\tableline
Injection, $Z=1$  index         & $\nu_{p}$   &  2.36 \\  
Injection, $Z>1$ index          & $\nu_{N}$   &  2.32 \\  
Transition rigidity             & $\R_{B}$ [GV] &    375  \\
Injection, slope-change         & $\Delta$    &   0.18 \\
Injection, low energy shaping   & $\eta_{s}$   & 0 \\
Injection, smoothing factor     & $s$          & 4   \\
Diffusion, normalization        & $K_{0}$ [kpc$^2$\,myr$^{-1}$]  & 0.026 \\
Diffusion, scaling index        & $\delta$       & 1/2  \\
Diffusion, low energy shaping   & $\eta_{t}$   &  -1 \\
Reacceleration, Alfv\'en speed  & $v_{a}$ [km\,s$^{-1}$] & 15  \\
Halo, half-height               & $L$ [kpc]        & 4  \\
Disk, half-height               & $h$ [kpc] & 0.1 \\
\tableline
\end{tabular}
\caption{Propagation parameter set for the reference model. \label{Tab::BenchmarkModelParameters}}
\end{center}
\end{table}
%
\begin{figure*}[!t]
\includegraphics[width=0.46\textwidth]{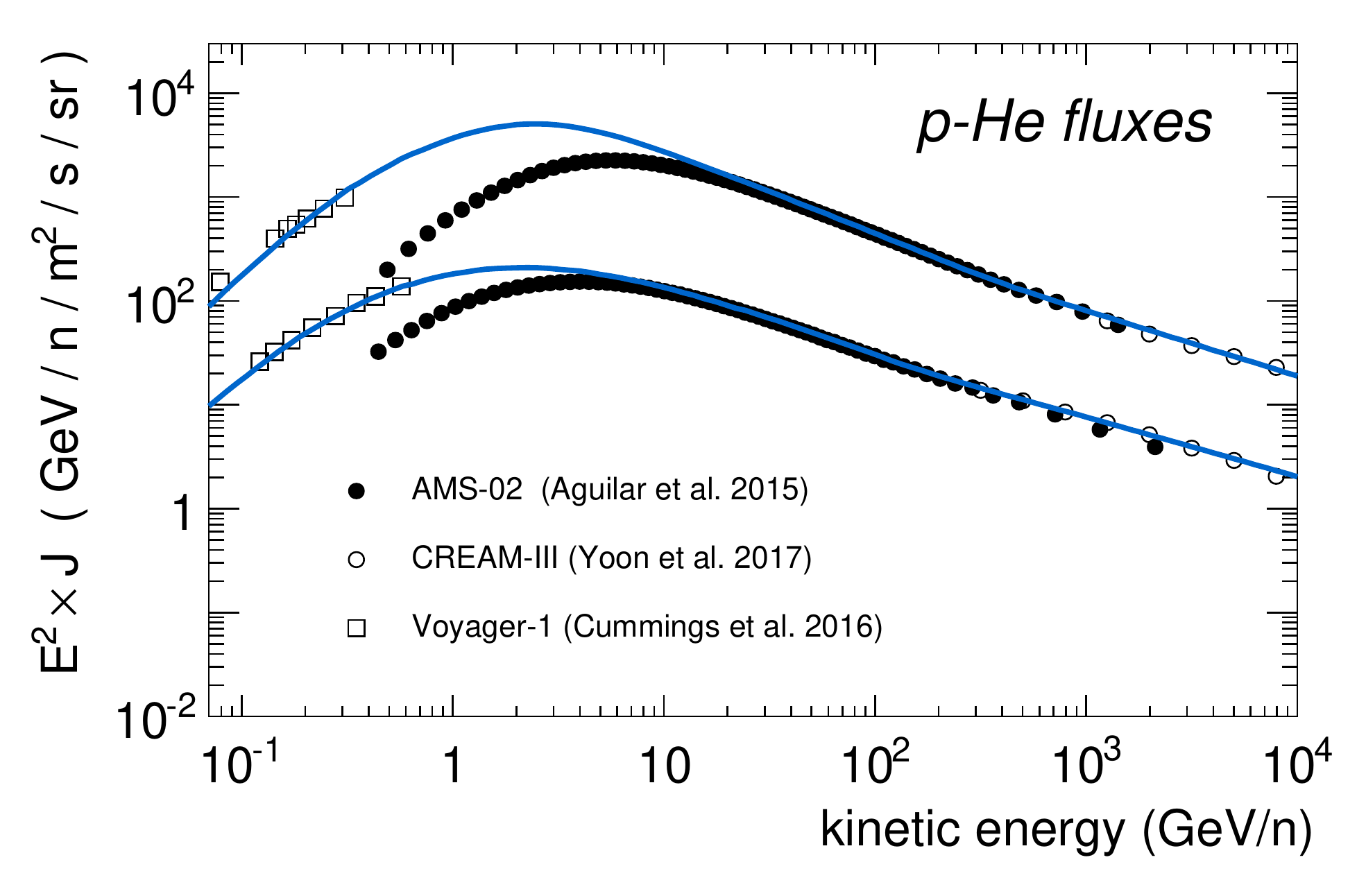} 
\quad
\includegraphics[width=0.46\textwidth]{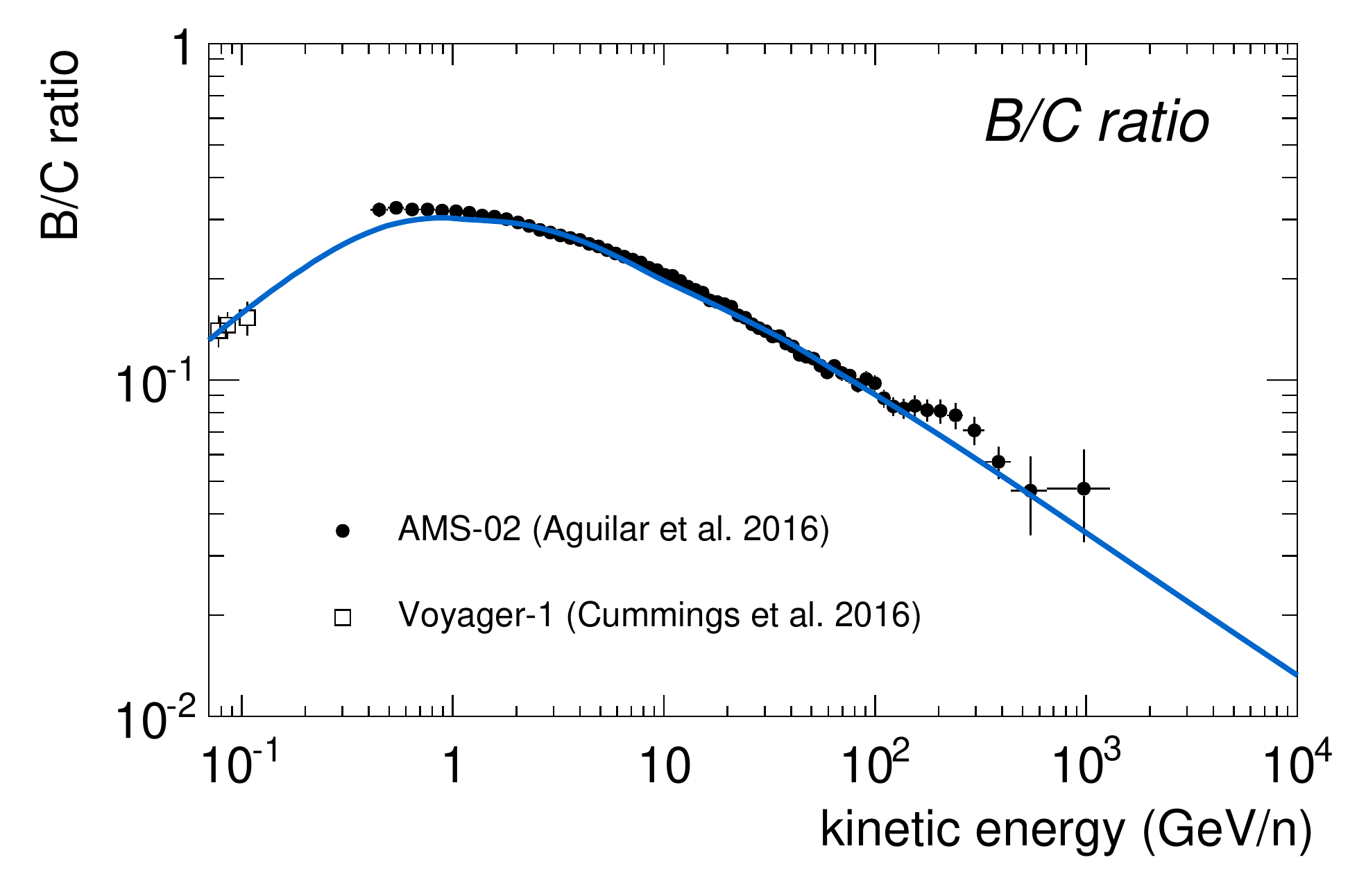} 
\caption{ 
  Reference model calculations for the proton and \He{} fluxes (left) and for the \BC{} ratio (right).
}\label{Fig::ccBenchmarkModel}
\end{figure*}
%
\begin{equation}\label{Eq::SourceSpectra}
\Q^{j}(\R) = q_{j}^{0}\beta^{\eta_{s}} \left(\R/\R_{0}\right)^{-\nu^{j}} \left[ 1 + \left(\R/\R_{\rm B}\right)^{s} \right]^{\Delta\nu/s} \,,
\end{equation}
where $\nu_{j}$ is particle dependent, and in particular is higher for proton, and $\Delta\nu$
describes a universal hardening occurring in all primary spectra at rigidity $\R_{B}$.
The parameter $\eta_{s}$ describes the behavior of the spectrum at low energy, with $\eta_{s}=0$ from the simplest linear diffusive-shock acceleration.
The composition factors $q_{j}^{0}$ are normalized to the primary CR abundances at reference rigidity $\R_{0}$.

The secondary production term $q_{j}^{sec}$ describes the products of decay and spallation of heavier CR progenitors.
For $k\rightarrow j$ decay of radioactive nuclei with corresponding lifetime $\tau_{0}^{k\rightarrow j}$, 
the source term is ${\Q}^{j}=\sum_{k} N^{k}({\bf r},E)/(\gamma\tau_{0}^{k\rightarrow j})$.
The occurrence of $k\rightarrow j$ fragmentation processes is described by:
\begin{equation}\label{Eq::GeneralSourceTerm}
{\Q}^{j}_{\rm sec}(E) = \sum_{k>j}  \int_{0}^{\infty} dE^{k} N^{k}(E^{k})  \sum_{i} \Gamma^{k\rightarrow j}_{i}(E,E^{k}) 
\end{equation}
where $\Gamma^{k\rightarrow j}_{i}$ is the differential rate of $j$-type particle production at kinetic energy $E$,
from collisions of $k$-type CR particles, with density $N^{k}$, with $i$-type targets of the ISM component (with number density $n_{i}$).
The ISM is essentially composed by hydrogen \Hyd{} and helium \He{} atoms, with $n_{H}\approx 0.9\,cm^{-3}$ and $n_{He}\approx 0.1\,cm^{-3}$.
Nuclear interactions couple the equation of each $j$-type nucleus to those of all heavier $k$-nuclei.
The system is resolved by starting from the heavier nucleus, which is assumed to be purely primary, and proceeding downward in mass.
The fragmentation loop is repeated twice.
The diffusion coefficient $K$ is taken as spatially homogeneous and rigidity dependent:
\begin{equation}\label{Eq::DiffusionCoefficient}
K(\R) = \beta^{\eta_{t}}K_{0}\left(\R/\R_{0}\right)^{\delta} \,,
\end{equation}
where $\K_{0}$ sets its normalization at reference rigidity $\R_{0}$, and the scaling index $\delta$ describes its rigidity dependence. 
The factor $\beta^{\eta_{t}}$ allows for a low rigidity change in the diffusion regime.
In particular, negative values for the parameter $\eta_{t}$ are used to to effectively account for a 
faster CR diffusion in the sub-GV rigidity region, which may be expected, \eg, from wave damping \citep{Ptuskin2006}.
This effect is reflected by the observed peak in the \BC{} ratio at $E\approx$\,1\,GeV/n. 
The characteristic shape of the \BC{} ratio in the GeV/n energy region, however, may be also ascribed to other processes such as
advection of strong diffusive reacceleration, if not to a change in spallation cross-sections \citep{Maurin2001,Jones2001}.
Reacceleration is described with a diffusion coefficient in rigidity space $\K_{RR}$ which is linked to spatial diffusion by the relation:
\begin{equation}
K_{RR}(\R)= \frac{4}{3}\;v_{a}^{2}\;\frac{\R^{2}/K(\R)}{\delta\,(4-\delta^2)\,(4-\delta)} \,.
\label{eq:Kpp}
\end{equation}
The parameter $v_{a}$ describes the average speed of Alfv\'en waves in the ISM medium.
We consider modest reacceleration ($v_{a}\cong$\,15\,km/s) under a scenario with Iroshikov-Kraichnan diffusion ($\delta=$1/2).
Other parameters are listed in Table\,\ref{Tab::BenchmarkModelParameters}.
The equation is solved in steady-state conditions $\partial \mathcal{N}_{j} / \partial t = 0$ at the boundary of the diffusion region.
The IS fluxes entering the solar system are then computed for each species:
\begin{equation}\label{Eq::FluxLIS}
J^{\rm IS}_{j}(E) = \frac{c A_{j}}{4{\pi}Z_{j}}\mathcal{N}_{j} \,,
\end{equation}
where $Z_{j}$ and $A_{j}$ are the CR charge and mass number, and the fluxes $J^{\rm IS}_{j}$ are given in units of kinetic energy per nucleon $E$.
The calculated ISs of proton and helium, along with the \BC{} ratio, are shown in Fig.\,\ref{Fig::ccBenchmarkModel} in comparison with new data from Voyager-1, \AMS, and CREAM-III.
The connection between model parameters and physics observables is discussed in several works \citep{Maurin2001,Maurin2010,Grenier2015}. 
In the purely diffusive regime, primary spectra behave as $J_{p}\sim\,hQ^{\rm pri}/\left( 2K/L + h\Gamma_{s} \right) \sim$\,$E^{-\nu-\delta}$,
while secondary-to-primary ratios follow $J_{s}/J_{p} \sim {\Gamma_{p \rightarrow s}}/( K/L + h\Gamma_{s} )$,
giving direct constraints to the parameters $\nu$, $K_{0}$, $L$, and $\delta$.
In the presence of energy changes, the connection between parameters and physics observables is blurred by other effects,
\eg, reacceleration, which significantly reshape the \BC{} ratio at $E\sim$\,0.1-50\,GeV/n.
To extract information in this energy region, our ability in modeling the unavoidable effect of solar modulation is then essential,
along with calculations of fragmentation reactions that lead to the production of boron nuclei.
The interpretation of primary spectra and secondary-to-primary ratios may be different in scenarios
where CR anomalies are ascribed to source components of \Li-\Be-\B{} nuclei,
nearby SNRs, or changes in diffusion\,\citep{TomassettiDonato2015,Blasi2017,Tomassetti2012Hardening,Maurin2010,Serpico2015}. 
In this work, however, these issues are not addressed.
The remainder of this paper is devoted to assess solar and nuclear physics uncertainties. 
For this task, the reference model present here will serve as the input model.

\begin{table*}[!t]
\begin{center}
\small
\begin{tabular}{ccccc}
\tableline
\tableline
NM station & NEWK & OULU &  APTY & JUNG  \\
\tableline
Detector type & 9-NM64 & 9-NM64 & 18-NM64 & 3-NM64\\
Location & Newark, Delaware  &  Oulu, Finland & Apatity, Russia & Jungfraujoch, Switzerland\\
Coordinates & 39.68\,N 75.75\,W & 65.05\,N, 25.47\,E & 67.55\,N, 33.33\,E & 46.55\,N, 7.98\,E\\
Altitude & 50\,m & 15\,m & 177\,m & 3570\,m\\
Cutoff & 2400\,MV  & 810\,MV  & 480\,MV & 4500\,MV \\
a  & (697.5 $\pm$ 4.5)\,MV   & (680.3 $\pm$ 4.8)\,MV & (709.8 $\pm$ 4.6)\,MV   & (689.9 $\pm$ 4.5)\,MV  \\
b  & 0.39 $\pm$ 0.01  & 0.40 $\pm$ 0.01  & 0.42 $\pm$ 0.01   & 0.38 $\pm$ 0.01  \\
$\chi^{2}$ & 2434.12 & 2393.75 & 2335.46 & 2426.4\\
\tableline
\end{tabular}
\caption{Fit results and properties of NM stations (from \url{http://www.nmdb.eu} \citep{Steigies2015}). \label{Tab::NMStations}}
\end{center}
\end{table*}

\section{Solar physics uncertainties}   
\label{Sec::SolarModulation}            
%
When entering the heliosphere, CRs travel across the outflowing solar wind embedded in its
turbulent magnetic field, where they undergo convection, diffusion, energy losses, and drift motion. 
To give a complete description of the solar modulation of CRs, all these effects have to be carefully modeled.
Our work takes advantage of recent progress in experimental measurements \citep{Bindi2017},
theoretical developments \citep{Potgieter2017}, and the availability of
numerical codes for solar modulation \citep{Kappl2016}.
%
\begin{figure*}[!t]
\includegraphics[width=0.82\textwidth]{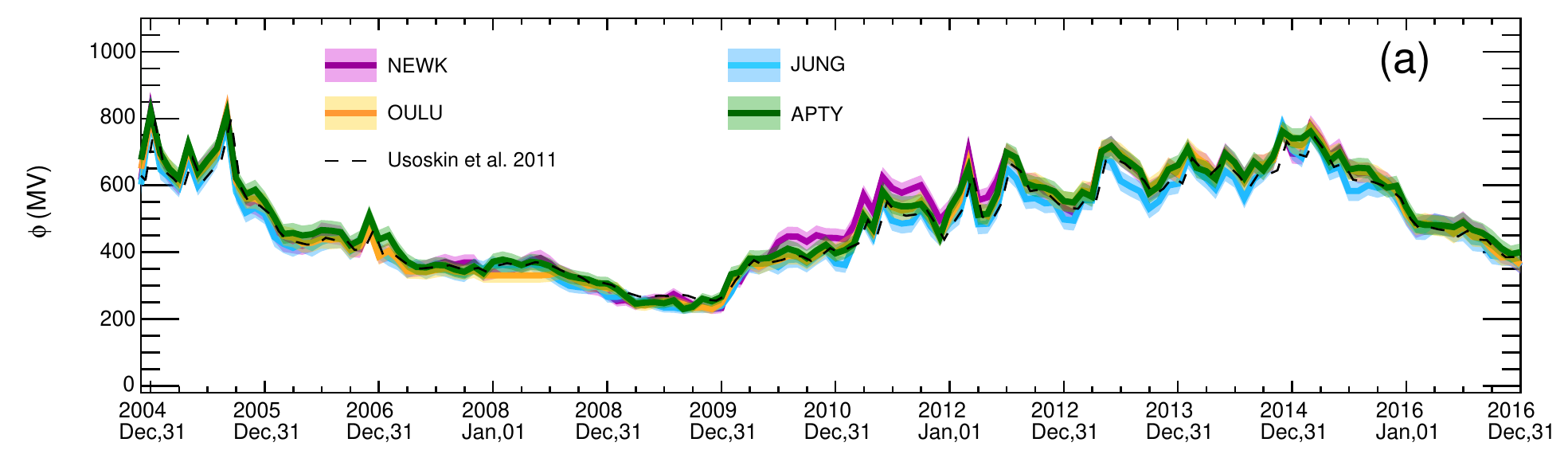} 
\quad 
\includegraphics[width=0.82\textwidth]{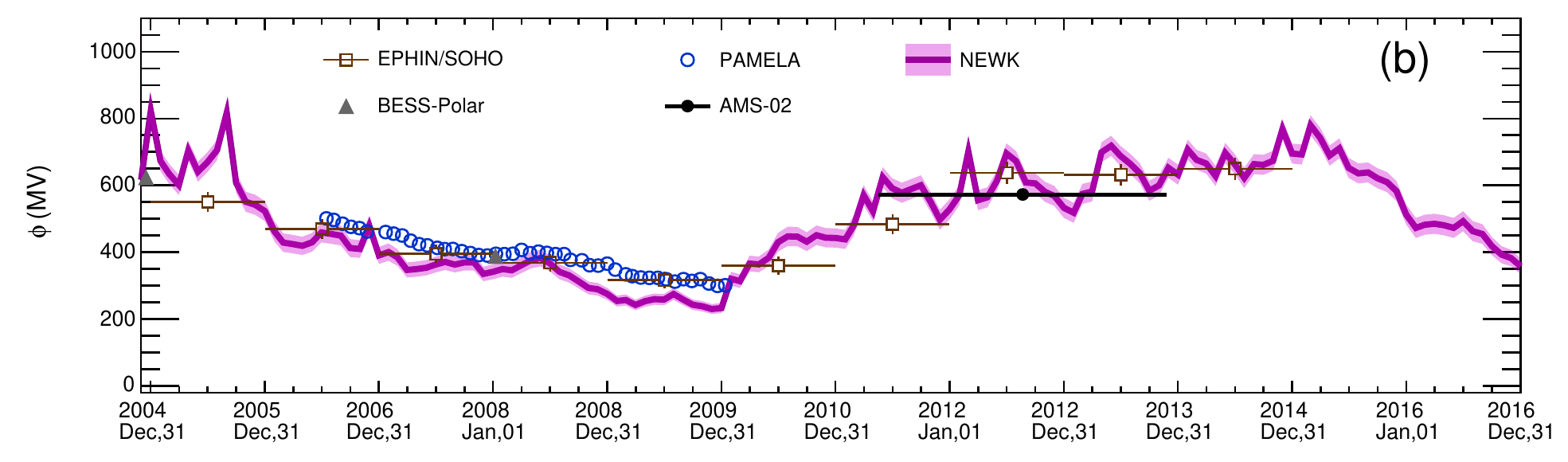} 
\quad 
\includegraphics[width=0.82\textwidth]{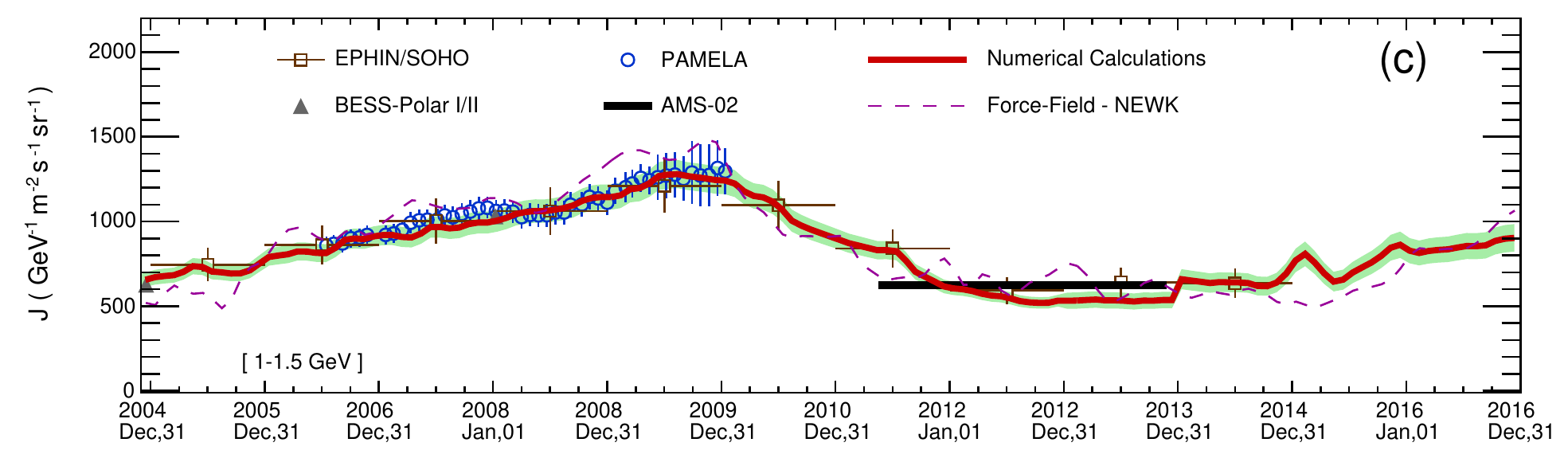} 
\caption{
  (a) Reconstruction of the modulation potential time-series $\phi$ using different NM stations. The thin dashed line is from Ref.\,\citep{Usoskin2011}.
  (b) Reconstruction of the modulation potential $\phi$ using direct fits to CR data in comparison with the $\phi_{\rm NEWK}$ time-series.
  (c) Time variation of the CR proton flux at $E=$1-1.5\,GeV. Calculations and uncertainties are shown in comparison with the data.
  The thin dashed line shows the force-field modulated flux using the $\phi_{\rm NEWK}$ time-series as the modulation parameter. 
}\label{Fig::ccProtonFluxTimeProfile}
\end{figure*}

\subsection{Solar modulation calculations}  
\label{Sec::SolarModulationCalculations}    
%
The goal of solar modulation calculations is to predict the modification of the CR energy spectrum $J^{\rm IS}(E)$
in heliosphere, at a given epoch $t$, via a transformation of the type:
\begin{equation}\label{Eq::Operator}
J(t,E) = \mathcal{\hat{G}}_{t}\left[J^{\rm IS}(E)\right] 
\end{equation}
The problem can be described in terms omnidirectional phase-space CR density $\psi=\psi({\bf r}, t)$, where the evolution is governed by the Krymsky-Parker equation:
\begin{equation}\label{Eq::Parker}
  \frac{\partial \psi}{\partial t}=
 {\bf \nabla}\cdot ( {\bf K} \cdot {\bf \nabla}\psi ) -( {\bf V}_{w} + {\bf V}_{d} ) \cdot {\bf \nabla}\psi + \frac{{\bf \nabla}\cdot {\bf V}_{w}}{3} \frac{\partial \psi}{\partial \ln \R}
\end{equation}
where the CR number density is given by $\mathcal{N} d\R=4\pi\R^{2}\psi$.
The various terms of Eq.\,\ref{Eq::Parker} represent convection with the solar wind, of speed ${\bf V}_{w}$, which is caused
by the expansion, particle drift motion with velocity ${\bf V}_{d}$, caused by gradients and curvature effects, 
spatial diffusion with tensor ${\bf K}$ representing its symmetric part, and energy losses due to adiabatic deceleration \citep{Potgieter2017}.
For each CR particle species, Eq.\,\ref{Eq::Parker} is set to zero in order to calculate steady-state solutions $\partial{\psi} /\partial t\equiv 0$,
which is a reasonable assumption for long-term modulation studies where the key parameters change gradually. 
The initial conditions are placed at the boundary of the modulation region,
where the reference model IS fluxes are provided from Eq.\,\ref{Eq::FluxLIS}.
A simplified approach that is commonly used in the CR astrophysics is the so-called \emph{force-field} (FF) approximation.
It consists in solving Parker's equation for a radially expanding wind, of speed 
$V_{w}(r)$, a fully isotropic diffusion coefficient with spatial-dependent part $K(r)$, after neglecting drift and loss terms. 
The FF approximation provides an analytical one-to-one correspondence between arrival fluxes at $r=r_{0}$ 
and IS fluxes at $r=r_{b}$ in terms of a lower rigidity shift, $\psi(r=r_{\rm max},\R+\phi)\cong \psi(r=r_{0},\R)$, where: 
\begin{equation}\label{Eq::PhiMeaning}
  \phi = \int_{r_{0}}^{r_{b}}  \frac{{V}_{w}(r)}{3 K(r)} dr
\end{equation}
The parameter $\phi$ is called the modulation potential. In terms of CR energy spectra $J(E)$, at a given epoch, 
kinetic energy per nucleon is shifted by $E=E^{\rm IS} - \frac{|Z|}{A} \phi$, so that the modulated flux is given by:
\begin{equation}\label{Eq::ForceField}
J(E) = \frac{(E+ m_{p})^{2}- m_{p}^{2}}{\left( E+m_{p} +\frac{|Z|}{A}\phi \right)^{2}-m_{p}^{2}} \times J^{\rm IS}(E + \frac{|Z|}{A}\phi)
\end{equation}
In this work, we do not make the FF approximation to compute the CR modulation, but we use quantity $\phi$ as input parameter.
We solve Eq.\,\ref{Eq::Parker} using a 2D numerical model with azimuthally symmetric spherical coordinates: 
radius $r$ and helio-colatitude $\theta$ \citep{Kappl2016}.
The solar wind is taken as radially flowing with speed $V_{w}\cong$\,400\,km\,s$^{-1}$.
The parallel component of the diffusion tensor is taken as $K_{\parallel}=  \kappa^{0} \frac{10^{22}\beta \R/{\rm GV}}{3B/B_{0}}$, 
in units of cm$^{2}$\,s$^{-1}$, and its perpendicular component is $K_{\perp}\cong 0.02\,K_{\parallel}$. 
The adimensional scaling factor $\kappa^{0}=\kappa^{0}(t)$  accounts for the time-dependence of the problem.
The regular magnetic field (HMF) $B$ follows the usual Parker model,
with $B=\frac{{A}B_{0}r_{0}^{2}}{r^{2}}\sqrt{1+\Gamma^{2}}$, where $\Gamma = \left(\Omega r/V_{w}\right) \sin{\theta}$, 
$\Omega=2.866\cdot\,10^{-6}$\,rad\,s$^{-1}$ is the Sun rotation speed, $B_{0}\cong$\,3.4\,nT sets the HMF at $r_{0}=$\,1\,AU,
and $A=\pm\,1$ sets the magnetic polarity state of the Sun.
The polarity is negative (positive) when the HMF points inward (outward) in the Northern Hemisphere.
This model accounts for drift effects, implemented as in \citep{Strauss2012,Kappl2016}.
Drift is important near the Heliospheric current sheet (HCS) where polarity changes from north to south.
When CR particles travel close to the HCS, their drift speed is a large fraction of 
their total speed, and 10$^{2-3}$ times larger than $V_{w}$. Away from the HCS, $V_{d}$ is of the order of $V_{w}$. 
The signs of its components depend on the product $qA$, which
is at the origin of the charge-sign dependence of CR modulation \citep{Strauss2012,Kappl2016}. 
Because of drift, particles and antiparticles sample different regions of heliosphere and their role interchange with polarity reversal. 
The angular extension of the HCS, which sets its waviness, is described by the tilt angle $\alpha$.
To obtain steady-state solutions of Eq.\,\ref{Eq::Parker} we employ a differential stochastic approach.
In particular we use the \SOLARPROP{} numerical engine \citep{Kappl2016}. 
With this method, Eq.\,\ref{Eq::Parker} is resolved by means of Monte Carlo generation of large samples of pseudoparticle trajectories. 
In practice, pseudoparticles are backward propagated from Earth to the modulation boundaries, that we set at the Heliopause HP, $r_{\rm hp}=\,122$\,AU. 
We also account for the termination shock (TS), placed at distance $r_{\rm ts}\cong$\,85\,AU \citep{Langner2003},
by smooth increase (decrease) of the HMF (wind speed) radial profiles
of a factor $s_{\rm ts}=$\,3 across the shock position, where $s_{\rm ts}$ is the TS compression ratio \citep{Potgieter2017}.
Due to the TS, CRs in the outer heliosphere follow slower diffusion and weaker convection \citep{Langner2003,Strauss2012}. 
The IS fluxes  $J^{\rm IS}$ , which state the boundary conditions at the HP,
are calculated within the reference model tuned against Voyager-1 and \AMS{} \citep{Cummings2016,Aguilar2015Proton}.
Further investigations on the phenomenology of this model will be presented in a future work.

\subsection{Uncertainties from solar modulation}  
\label{Sec::SolarModulationUncertainties}         
%
%
\begin{figure}[!th]
\begin{center}
  \includegraphics[width=0.48\textwidth]{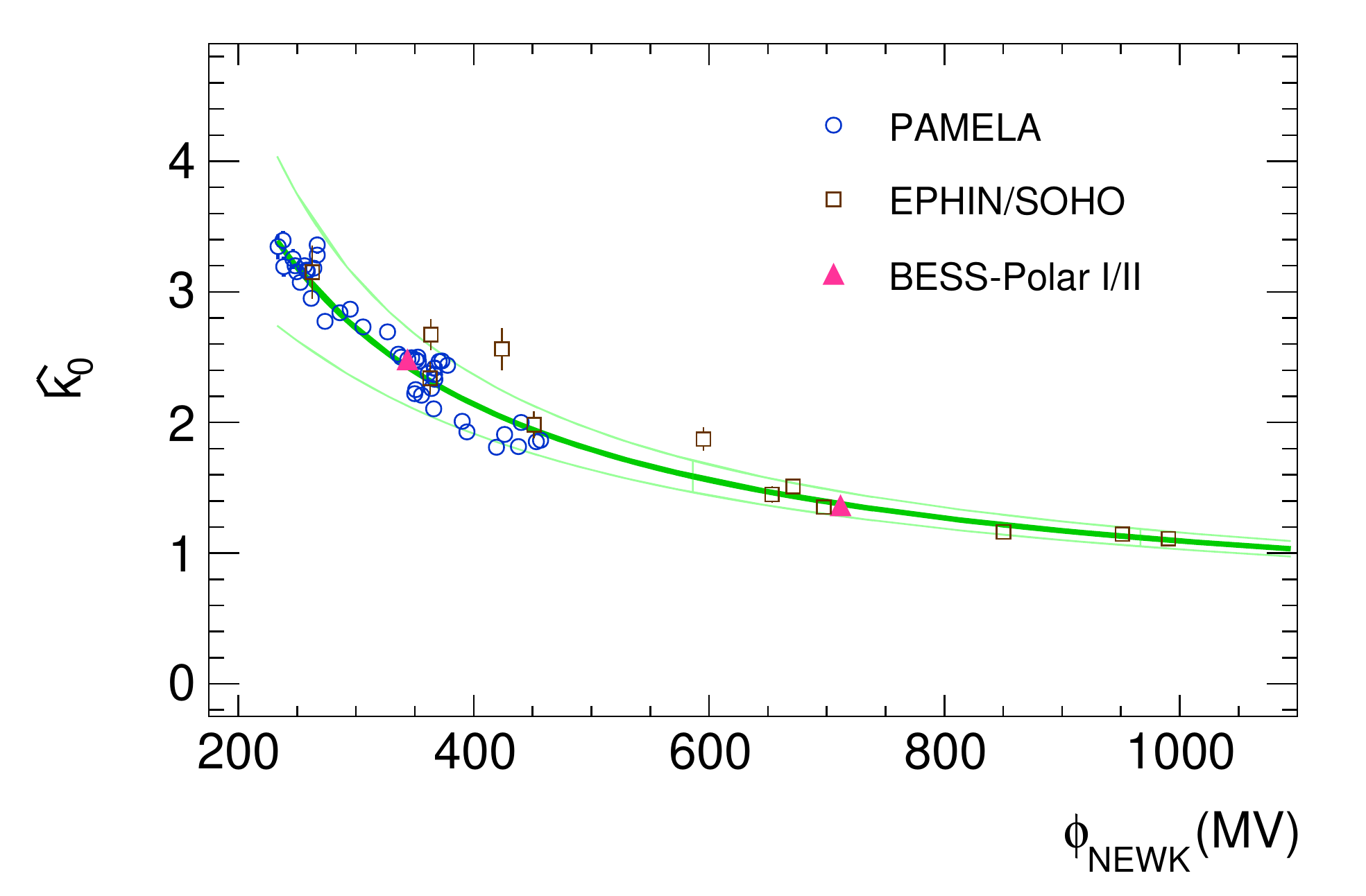} 
\caption{%
  Best fit of the diffusion coefficient scaling factor $\hat{\kappa}^{0}$ obtained from various CR data sets and 
  plotted against the corresponding modulation potential, $\phi$, reconstructed using NEWK data.
  \label{Fig::ccKAPPAFitVSPHI}
}
\end{center}
\end{figure}
%
%
\begin{figure*}[!t]
\includegraphics[width=0.92\textwidth]{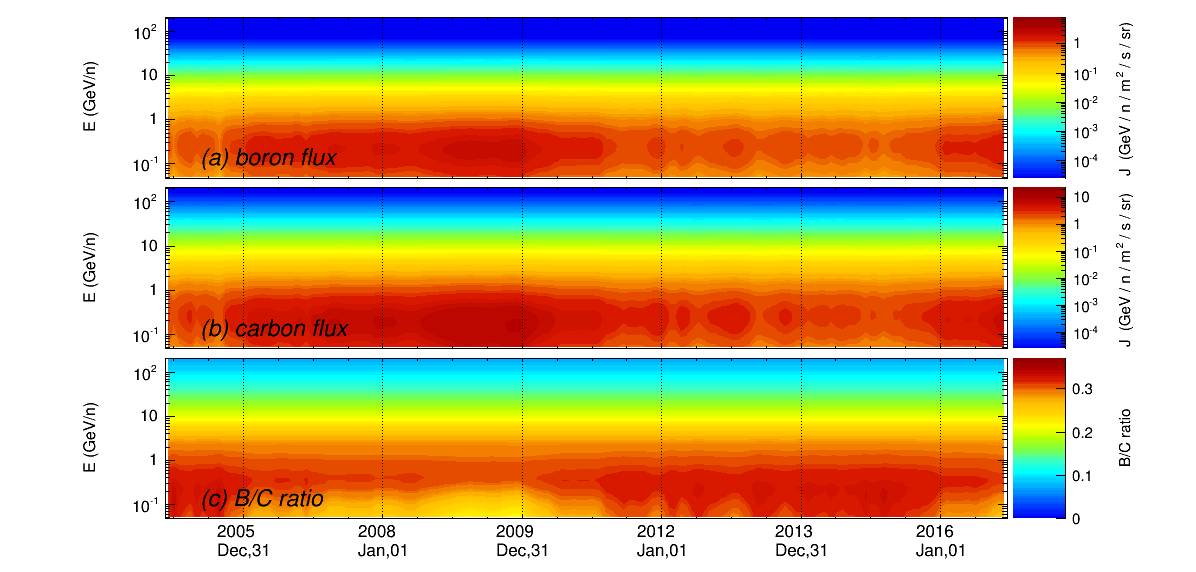}
\caption{ 
  Calculated temporal variation of (a) boron and (b) carbon energy spectra, along with (c) the \BC{} ratio between 2005 and 2016.
  The flux intensities and the ratio values are represented by the color bars.
}\label{Fig::ccNucleiFluxVariation2D}
\end{figure*}
%
To assess the uncertainties on the \BC{} calculations that arise from solar modulation modeling,
we perform a global data-driven reconstruction of the time evolution of the CR flux based on NM counting rates and CR proton data.
With this reconstruction, we constrain the key input parameters of the solar modulation model and their time dependence,
which allows us to estimate their influence on the predicted time evolution of the \BC{} ratio.
This procedure provides a time-dependent uncertainty band for our model prediction, calculated on monthly basis,
which can therefore be averaged over a the period of the \AMS{} observation time.

From the implementation described in Sec.\,\ref{Sec::SolarModulationCalculations}, two time-dependent parameters have to be determined:
the normalization of the diffusion coefficient, $\kappa_{0}=\kappa_{0}(t)$, and the tilt angle of the HCS, $\alpha=\alpha(t)$.
To model the tilt angle, we adopt a smooth interpolation of the time-series $\alpha(t)$ based on the ``radial model'' reconstruction \citep{Hoeksema1995}.
This reconstruction is provided on 10-day basis by the \emph{Wilcox Solar Observatory} \citep{WSO}. 
To determine the time evolution of the diffusion coefficient normalization, we make use of the monthly resolved series of NM rates.
For this task, we convert monthly average NM counting rates into a time-series of modulation potential $\phi=\phi(t)$ that, as discussed, arise from the FF solution of Eq.\,\ref{Eq::Parker}.
An inverse relationship between $\phi$ and $\kappa_{0}$ is suggested by Eq.\,\ref{Eq::PhiMeaning}, from which $\phi\sim\,V/\kappa_{0}$.
Here we simply write $k_{0}(t)\equiv a\phi^{-1}_{d}(t) + b$, where the coefficients $a$ and $b$ are free parameters
that we determine using CR flux data, and $\phi_{d}$ is determined using NM counting rates. 
We recall that, although we make use of the parameter $\phi$ obtained in the context
of the FF approximation, we do not make the FF approximation in our flux calculations.

The procedure to determine the time-series $\phi_{d}(t)$ follows early works \citep{Tomassetti2017PHe,Ghelfi2017}.
For a given NM detector $d$, located altitude $h_{d}$ and geomagnetic cutoff $\R_{C}^{d}$,
the link between the counting rate $\mathcal{R}_{\rm NM}^{d}$ and top-of-atmosphere CR fluxes $J_{j}$ (with $j=$\p,\,\He) is expressed by:
\begin{equation}\label{Eq::NMRate}
  \mathcal{R}_{\rm NM}^{d}(t) = \int_{0}^{\infty}  dE \cdot \sum_{j={\rm CRs}} \mathcal{H}^{d}_{j}(E)\cdot\mathcal{Y}^{d}_{j}(t,E)\cdot J_{j}(t,E)
\end{equation}
where $\mathcal{H}^{d}$ is a \emph{transmission function} around the cutoff value $\R^{d}_{C}$, parametrized as a smoothed step function \citep{SmartShea2005,Tomassetti2015XS},
and $\mathcal{Y}^{d}_{j}$ is the $j$-particle dependent detector response function \citep{Maurin2015NM}.
In particular we use a factorized form, $\mathcal{Y}^{d}_{j} = \mathcal{V}^{d} \mathcal{F}^{d}_{j}$,
where $\mathcal{F}^{d}_{j}(t,E)$ accounts for time and energy dependencies of the NM response, including the development of hadronic cascades \citep{Cheminet2013},
and the factor $\mathcal{V}^{d} \propto exp(f_{d}h_{d})$ set the absolute normalization and its altitude dependence. 
To compute Eq.\,\ref{Eq::NMRate}, 
Equation\,\ref{Eq::NMRate} is calculated using monthly averaged NM rates, corrected for detector efficiency and pressure, 
provided between January 2005 and January 2017 from various stations \citep{Steigies2015}. 
The NM detectors considered in this works are listed in Table\,\ref{Tab::NMStations}.
For a given station $d$, the parameter $\phi$ is obtained from the request of agreement between
the calculated rate $\mathcal{{R}}^{d}$ and the observed rate $\hat{\mathcal{R}}^{d}$,
together with the requirement that $\int_{\Delta T^{d}}{R}^{d}(t) dt = \int_{\Delta T^{d}} \hat{R}^{d}(t) dt$
where the integration is performed over the observation periods $\Delta T^{d}$ \citep{Tomassetti2017PHe}.
For all detectors we consider $\Delta{T^{d}}=$12 years, between January 2005 and January 2017.
The time-series of monthly reconstructed $\phi$ resulting from this procedure is shown in 
Fig.\,\ref{Fig::ccProtonFluxTimeProfile}a using the NM detectors of Table\,\ref{Tab::NMStations}. 
Uncertainty bands are shown for each station, corresponding to $\delta\phi\sim$\,25\,MV \citep{Ghelfi2017}. 
This is also the level of discrepancies among the various NM responses.
The $\phi(t)$-series calculated from Ref.\,\citep{Usoskin2011} is superimposed as dashed lines, for reference. 
This time-series is in agreement with our reconstruction. 

From the reconstructed time-series of modulation potential,
the parameters $a$ and $b$ that specify the time-evolution of the diffusion scaling $\kappa_{0}(t)$.
are determined using CR proton data from PAMELA, EPHIN/SOHO, and BESS-Polar, for a total of to 3993 CR data points.
Using CR flux measurements  $\hat{J}_{i,k}=\hat{J}(t_{i},E_{k})$, collected at given epoch $t_{i}$ and energy $E_{k}$,
we build a global $\chi^{2}$ estimator:
\begin{equation}\label{Eq::GlobalChiSquare}
\chi^{2}(a,b) = \sum_{i,\,k} \left[ \frac{ J(t_{i}, E_{k}; a, b) - \hat{J}_{i,k} }{\hat{\sigma}_{i,k} } \right]^{2} 
\end{equation}
The proton flux calculations $J(t_{i}, E_{k})$ are carried out from numerical solutions of Eq.\,\ref{Eq::Parker}
using input parameters $\alpha_{i}=\alpha(t_{i})$ and $\kappa_{i}^{0}= a\phi^{-1}_{i} + b$. Both time-series $\alpha_{i}$ and $\phi_{i}$
have been smoothly interpolated with monthly resolution.

The best fit parameters $\hat{a}$ and $\hat{b}$ are obtained using standard $\chi^{2}$-minimization techniques.
The minimization required a large scan over the $\kappa^{0}$-$\alpha$ parameter space,
for a total simulation of $\sim$\,1.9 billion particle trajectories.
This procedure provides the time and energy dependence of the proton flux and its associated uncertainties.
The time profile of the CR proton flux at $E\approx\,1$\,GeV is shown in Fig.\,\ref{Fig::ccProtonFluxTimeProfile}b, in comparison with the data. 
The procedure was repeated using the four time-series of $\phi_{d}$, giving slightly different --but consistent-- best fit parameters.
The best fit results are listed in Table\,\ref{Tab::NMStations} for the various NM detectors considered.
Nonetheless, the robustness of the predictions is mostly related to the constraints provided by the CR proton flux data,
which is plotted as shaded band in Fig.\,\ref{Fig::ccProtonFluxTimeProfile}b.
From the figure, one can also  notice a clear anticorrelation between CR flux and modulation potential.
The modulated fluxes presented in the figure are calculated on monthly based between 2005 and 2017, therefore covering 
a 11-year period corresponding to a complete solar cycle. This period also includes magnetic reversal, which occurred in early 2013,
where the Sun's polarity state switched from $A<0$ to $A>0$, shown in the figure as vertical shaded bar.
It can be seen that the calculations agree well with the \AMS{} proton data collected between May 2011 and November 2013,
although these data have not been included in the $\chi^{2}$ estimator.
We also show, in Fig.\,\ref{Fig::ccKAPPAFitVSPHI}, the inferred diffusion scaling $\hat{\kappa}^{0}_{i}$ 
obtained by performing single fits to the CR energy spectra plotted against the modulation potential $\phi_{i}^{\rm NEWK}$ reconstructed 
at the same epoch using data from the NEWK station. 
The solid line represents the relation $\kappa^{0}=\hat{a}\phi_{d}^{-1}+\hat{b}$, for the NEWK station, and its uncertainty band.
This figure shows the inverse relationship between that the two quantities.
Once the parameters are constrained, the model can be used to predict CR flux of other species including heavier nuclei or antiparticles. 
In Fig.\,\ref{Fig::ccNucleiFluxVariation2D} we plot the time evolution of the energy spectra of boron (a) and carbon (b), along with their ratio (c)
calculated between 2005 and 2017. The figure represents the flux (ratio) values as a function of energy and time.
In our calculations for the \BC{} ratio, we have accounted for various sources of model uncertainties, including
uncertainty on the input IS fluxes, uncertainty on the input parameters and on the global fitting procedure.
The resulting error $\delta_{\rm B/C}$ is mildly time-dependent. 
The resulting uncertainties on the \BC{} ratio are presented in Sec.\,\ref{Sec::Results} after performing a time average over the \AMS{} observation period.

\subsection{On the force-field approximation}  
\label{Sec::FFApproximation}                   
%
\begin{figure}[!th] 
\begin{center}
  \includegraphics[width=0.48\textwidth]{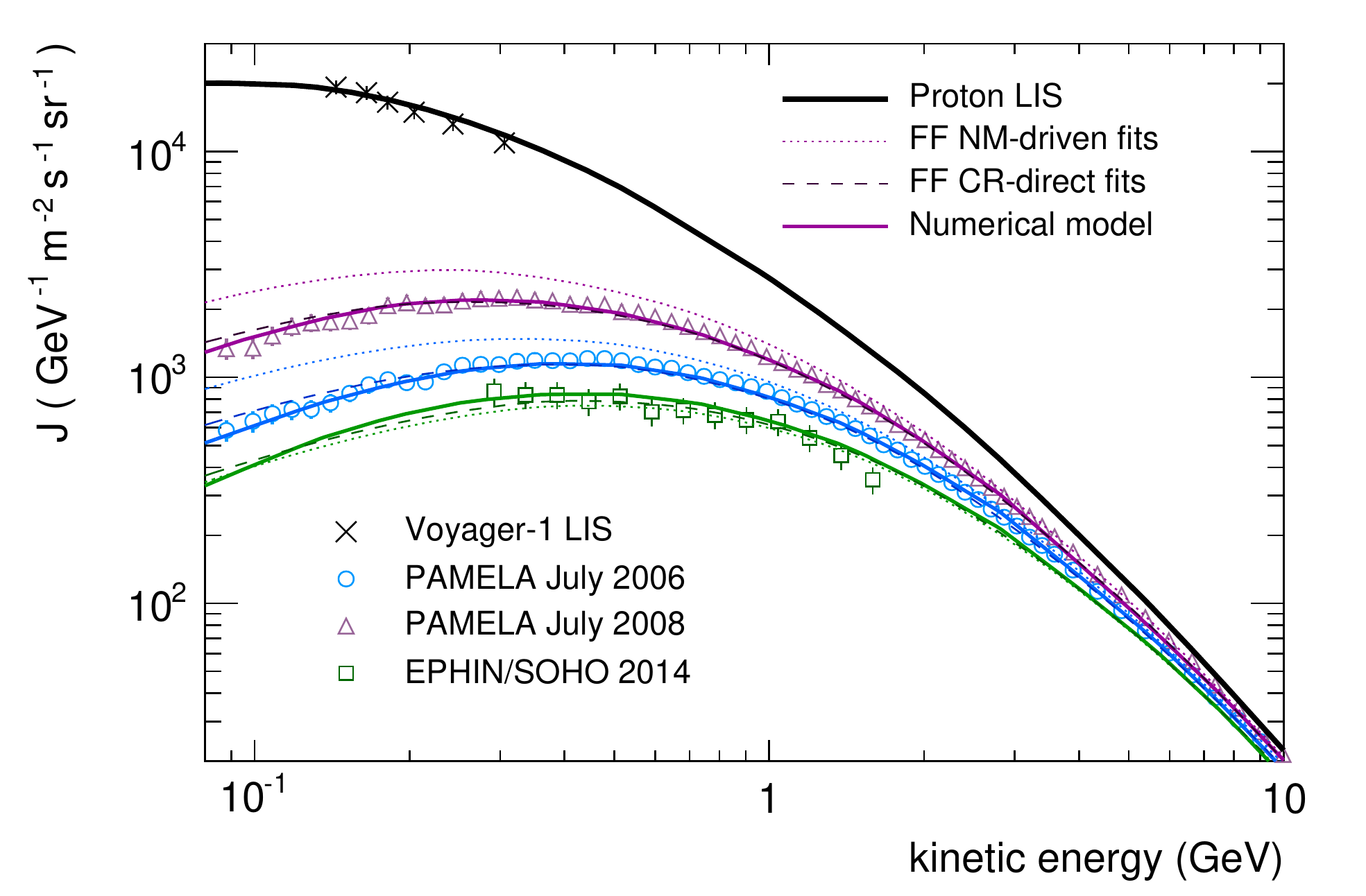}
\caption{%
  Measurements of the CR proton spectrum at different epochs in comparison with
  FF calculations with NM-driven (dotted lines) and CR-driven (dashed lines) modulation parameters,
  and from the numerical model of Sec.\,\ref{Sec::SolarModulationCalculations}. 
  \label{Fig::ccProtonFluxVSEkn_NMvsCRvsNC}
}
\end{center}
\end{figure}
%
In our calculations, we made use of the modulation parameter $\phi$ as input. This parameter is obtained in the context
of the FF approximation, although we adopted a numerical approach to solve the Krymsky-Parker equation.
On may wonder if a fully FF-based approach may reach an adequate level of accuracy.
In Fig.\,\ref{Fig::ccProtonFluxTimeProfile}b, \emph{NM-driven} FF modulation potential derived from NEWK
is compared with the $\phi$-values determined from direct fits to CR spectra (\emph{CR-driven}) of the various data sets.
While the overall time profile is well reproduced in both approach, some inconsistencies are apparent, \eg, in the PAMELA data.
Symmetrically, in Fig.\,\ref{Fig::ccProtonFluxTimeProfile}c, numerical calculations (thick line) are shown together the
NM-driven FF calculations (thin dashed line) obtained using the $\phi_{\rm NEWK}$ parameter of Fig.\,\ref{Fig::ccProtonFluxTimeProfile}a
to FF-modulate the proton IS. Again, some inconsistency can be observed with the PAMELA data.
Similar tensions were also noted in other studies \citep{Cholis2016}.
The origin of this tension is that, with the integration of the NM rates of Eq.\,\ref{Eq::NMRate}, the modulation parameter 
is determined at the $\mathcal{O}$(10\,GeV) energy scale. In contrast, with the CR-driven approach, the fit
is performed over the whole energy spectra, \ie, down to 80\,MeV for PAMELA, which is where the FF approximation breaks down.
More in general, the larger discrepancies are expected during $q{A}<0$ periods (when drift is relevant)
and especially at low energy or during high level of solar activity (where the modulation effect is stronger).
Nonetheless, when the FF is directly applied for fitting CR data (with $\phi$ as free parameter) the flux calculations can be described reasonably well.
This is illustrated in Fig.\,\ref{Fig::ccProtonFluxVSEkn_NMvsCRvsNC}. Along with the IS proton flux, solar modulated
spectra are shown from numerical calculations, from NM-driven FF calculations (with $\phi$ determined from NEWK data),
and from CR-driven FF calculations (where the flux is directly fit to the CR data of the figure). 
The CR-driven approach leads to a fairly good description of the data, within
10-15\,\% of precision for the fluxes of Fig.\,\ref{Fig::ccProtonFluxVSEkn_NMvsCRvsNC}.
The main weakness of this approach lies in its limited predictive power.
For instance, FF-calculations calibrated with proton data cannot be used to describe antiprotons, which require different $\phi$-values.
Similarly, FF-calculations calibrated with NM data under a given polarity state cannot be applied to the opposite polarity. 
Hence the FF method can provide an adequate description of the CR spectra, but it cannot be used in a predictive way,
\eg, for estimating the astrophysical background of CR antiparticles.
Having a predictive model is also necessary to describe the time-unresolved collected over long exposures.
For instance, the \AMS{} data represent a time average over years of observation time. 
The CR flux evolves during this period, which is only covered by NMs, on much shorter timescales. 
To account for these issues of the simple FF method, recent works proposed easy-to-use generalizations of
the FF formula through the introduction of a charge- or rigidity-dependent parameter $\phi$ \citep{Cholis2016}.

\section{Nuclear physics uncertainties}  
\label{Sec::FragmentationCrossSections}  
%
The determination of the CR transport parameters using the \BC{} ratio relies obviously on calculations
of the boron production rate from the disintegration or decay of the heavier nuclei.
The accuracy of secondary production calculations therefore depends on the
reliability of the production and destruction cross-sections employed.

\subsection{Cross-section calculations} 
\label{Sec::CrossSectionCalculations}   
\begin{figure*}[!thb]
\begin{center}
  \includegraphics[width=0.75\textwidth]{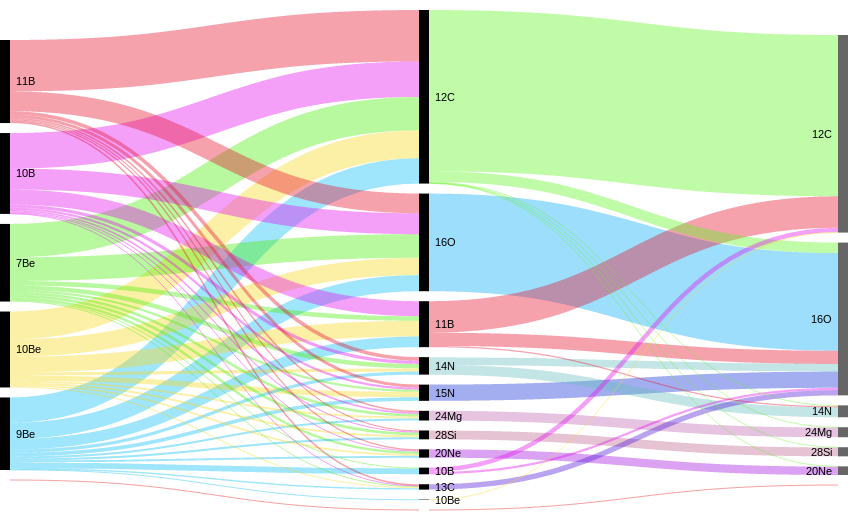}
\caption{ 
  Alluvial diagram of nuclear fragmentation contributions 
  from primary CR elements (right blocks) to the abundance of \Be{} and \B{} isotopes (left blocks) at $E=$\,1\,GeV/n of kinetic energy.
  Two steps are shown, illustrating the role of intermediate long lived nuclei.
  \label{Fig::ccNuclearFragmentationAlluvialPlot}
}
\end{center}
\end{figure*}

%
The general source term of secondary production of CR nuclei is given in Eq.\,\ref{Eq::GeneralSourceTerm}.
This expression is further simplified by the \emph{straight-ahead} approximation,
according to which it the secondary nucleus is assumed to be ejected with the same kinetic energy per nucleon of the
fragmenting projectile, \ie, $d\sigma/dE(E,E^{\prime})\cong\sigma(E)\delta(E-E^{\prime})$.
Furthermore, the composition of the interstellar gas is dominated by hydrogen (90\,\%) and helium (10\,\%), 
so that reactions involving heavier target can be safely neglected. Hence the source term for $j$-type particle production becomes:
\begin{equation}\label{Eq::GeneralSourceTerm}
{\Q}^{j}_{\rm sec}(E) = \sum_{k>j} c\beta_{k}  N^{k}(E) \left[ n_{\Hyd}\sigma_{\Hyd}^{k\rightarrow{j}}(E) + n_{\He}\sigma_{\He}^{k\rightarrow{j}}(E) \right]
\end{equation}
In CR propagation models, semiempirical formulae are used to predict the \XSs{} $\sigma^{\rm{P}\rightarrow\rm{F}}$ 
at any energy and for all relevant \textit{projectile}$\rightarrow$\textit{fragment} (\PF) combinations.
These algorithms are based on observed systematics in mass yields, charge dispersion, and energy dependence of the fragments produced in nucleus-nucleus collisions.
Popular algorithms are \YIELDX{} \citep{Silberberg1998}  and \WNEW{} \citep{Webber1990WNEW,Webber2003}.
The \GALPROP{} code makes use of parametric formulae obtained from fits to the data or from nuclear
codes \texttt{CEM2k} and \texttt{LAQGSM},  eventually normalized to the data \citep{StrongMoskalenko1998,MoskalenkoMashnik2003}. 
Along with mass-changing and charge-changing reactions, \XSs{} for isotopically separated targets or
fragments have been measured by several experiments (see Sec.\ref{Sec::Observations}). 
In spite of such large collection of \PF{} reactions, the data are available only at $\sim$\,0.1\,5\,GeV/n energies.
In this work, we mostly are interested in the production of $^{10}$\B, $^{11}$\B, $^{7}$\Be, $^{9}$\Be, and $^{10}$\Be{} isotopes
from interactions of $^{10,11}$\B, $^{12}$\C, $^{14,15}$\N{} and $^{16}$\Oxy{} off hydrogen and helium targets.
These reactions  account for $\gtrsim$\,90\% of the \Be{} and \B{} production. The remaining
reactions involve fragmentation of heavier nuclei such as $^{10}$\Ne, $^{28}$\Si, or $^{56}$\Fe, which give
a minor contribution to the boron production uncertainty.
It is important to stress that the fragmentation network of CRs in the Galaxy may involve several steps
where the disintegration or decay of intermediate particles contributes to the abundance of a given final-state CR nucleus.
If the intermediate nucleus is stable or long lived (\ie, its lifetime is larger in comparison with the CR propagation timescale)
this species has to be properly accounted in the CR propagation network. 
Examples of this kind are $^{16}$\Oxy$\rightarrow$$^{12}$\C$\rightarrow$$^{11}$\B{} or $^{16}$\Oxy$\rightarrow$$^{15}$\N$\rightarrow$$^{11}$\B{} reaction processes.
The multistep nature of CR fragmentation is illustrated in the alluvial diagram of Fig.\,\ref{Fig::ccNuclearFragmentationAlluvialPlot} 
which links the secondary \Be-\B{} isotopes (left blocks) to their purely primary progenitors (right blocks)
via two stages of fragmentation calculated at 1\,GeV/n of kinetic energy.
The stream fields between the blocks represent the fragmentation reaction channels (to be read from right to left) 
sized accordingly to their contribution to secondary abundances.
From this graph it can be seen, for example, that the $^{10}$\B{} isotope is mostly generated from $^{12}$\C{} and $^{16}$\Oxy{}
as expected, while a non-negligible contribution is represented by fragmentation of secondary species such as $^{11}$\B, $^{15}$\N, or $^{13}$\C.
Thus, the usual view of purely primary CRs fragmenting into secondaries is only an idealized approximation.
The actual calculations account for a complex multistep reaction network to determine the abundance of \Li-\Be-\B{} particles.

\begin{figure}[!th]
\begin{center}
  \includegraphics[width=0.48\textwidth]{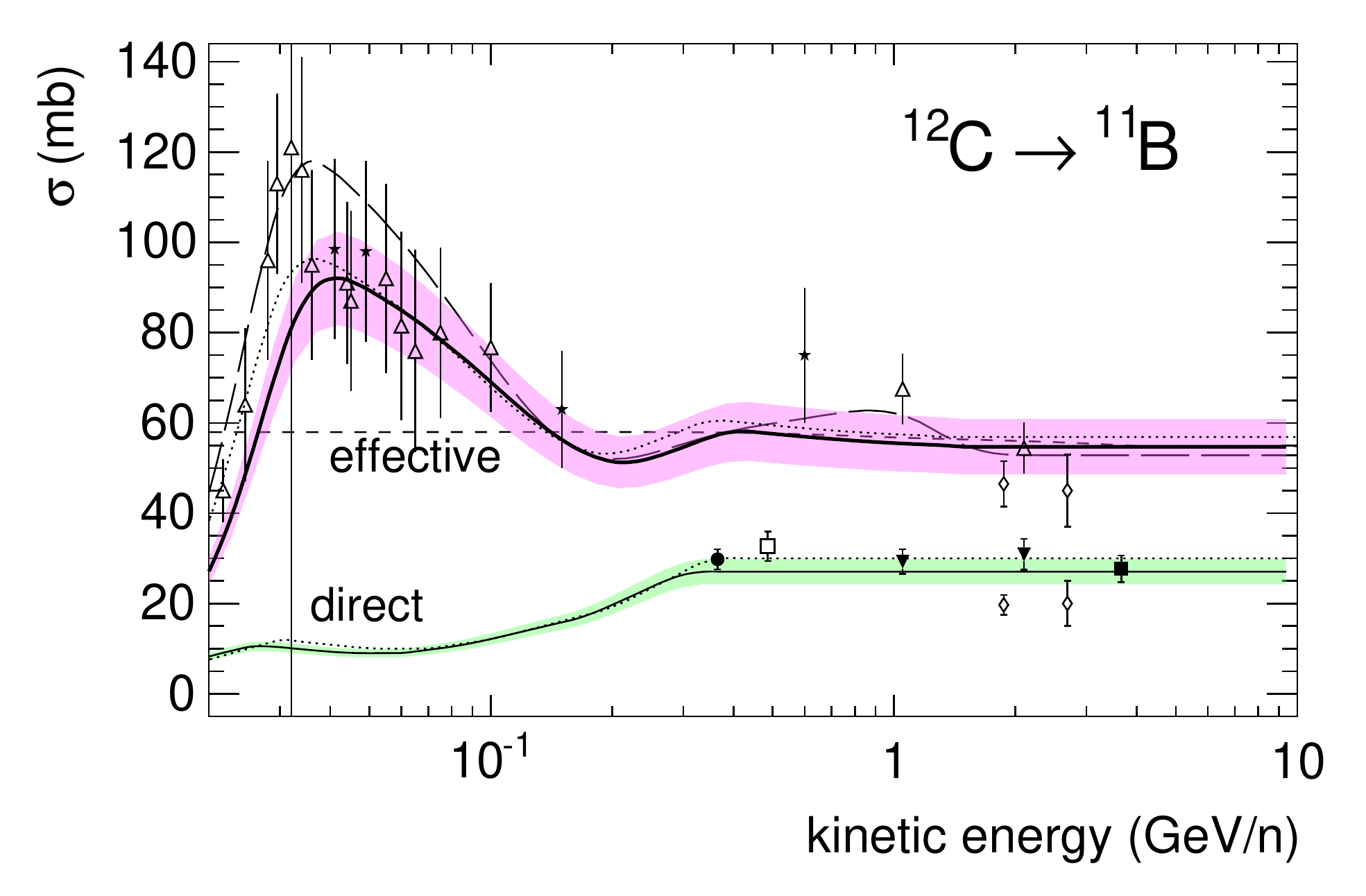}
\caption{%
Cross-section measurements and calculations for the production of $^{11}$\B{} isotopes from fragmentation of $^{12}$\C{} off  hydrogen target. 
The direct $^{12}$\C$\rightarrow$$^{11}$\B{} channel (green line) is shown together with the cumulative reaction (purple line). 
The latter includes the process $^{12}$\C$\rightarrow$$^{11}$\C$\rightarrow$$^{11}$\B{} mediated by the ghost nucleus $^{11}$\C.
The data symbols are given in Sec.\ref{Sec::Observations}. 
\label{Fig::ccGhostExampleC12toB11}
}
\end{center}
\end{figure}

Another situation is when the intermediate nucleus is unstable with a short lifetime to the CR propagation timescale.
An example is given by $^{11}$\B{} particles that can be generated either by ``direct'' one-step processes
(\eg, $^{12}$\C$\rightarrow$$^{11}$\B{}) or from production of intermediate isotopes (\eg, $^{11}$\C) with subsequent decay into $^{11}$\B. 
In all these cases, intermediate short-lived nuclei are  treated as ``ghost'' so that they are accounted using effective \XS{} formulae:
\begin{equation}\label{Eq::GhostNuclei}
\sigma^{k\rightarrow{j}}_{\rm eff} \equiv \sigma_{\rm direct}^{k\rightarrow{j}} + \sum_{g} \sigma^{k\rightarrow{g}} \mathcal{B}^{g\rightarrow{j}} \,,
\end{equation} 
where $\mathcal{B}^{g\rightarrow{j}}$ is the branching ratio for the $g\rightarrow{j}$ decay channel \citep{Maurin2001}. 
Ghost nuclei do not experience CR propagation within their lifetime, but they can be detected
in laboratory measurements, depending on the time-resolution of the experiment.
An instructive example is given in Fig.\,\ref{Fig::ccGhostExampleC12toB11} for the $^{12}$\C$\rightarrow$$^{11}$\B{} reaction channel.
In this reaction, measurements and calculations are shown for the ``direct'' $^{11}$\B{} production (green line) 
and for the cumulative production (purple line) which includes $^{11}$\C{} ghost nuclei.
Experimentally, the latter is measured by the sole mass identification of $A=11$ end-products.
This processes is the relevant one for CR propagation and represents the effective \XS{} of Eq.\,\ref{Eq::GhostNuclei}.
All fragmentation \XSs{} used in this work have to be meant as effective ones.
Finally, to properly compute the abundance of boron at the GeV/n energy scale, it is also important to model the
production of beryllium isotopes via iterative calculations. Beryllium and boron share the same progenitors, 
but the small component of $^{10}$\Be{} contained in it the \Be{} flux decays radioactively into $^{10}$\B, 
contributing to nearly 10\,\% in the \BC{} ratio at low energy.

\subsection{Uncertainties from nuclear fragmentation}  
\label{Sec::CrossSectionUncertainties}                 
%
We utilize a large compilation of \XS{} data on isotopically resolved reactions. An extensive
survey of the literature can be found in other works \citep{Tomassetti2015XS,MoskalenkoMashnik2003,Moskalenko2013}.
To determine the \XS{} uncertainties using the data, we perform a data-driven renormalization of the \GALPROP{} parametrizations $\sigma_{\rm G}(E)$.
Our approach follows the procedure of earlier studies on \Hyd{}\,--\,\He{} isotopes \citep{Tomassetti2012Isotopes}
and heavy nuclei \citep{Moskalenko2013,Tomassetti2015XS}, although we introduce some improvements.
For all \PF{} channels, parametric \XS{} formulae have been redetermined via minimization of the following quantity:
\begin{equation}\label{Eq::XSRefit}
\chi^{2}_{{\rm P}\rightarrow{\rm F}} = \sum_{i} \left[ \frac{ \hat{\sigma}^{i}_{{\rm P}\rightarrow{\rm F}} - \xi\cdot \sigma_{{\rm P}\rightarrow{\rm F}}(\eta\cdot E) }{\delta^{i}_{{\rm P}\rightarrow{\rm F}}} \right]^{2}
\end{equation}
where $\delta^{i}_{{\rm P}\rightarrow{\rm F} }$ is the experimental error on the $i$-th data point for the considered \PF{} channel.
%
%
\setlength{\tabcolsep}{0.034in} 
\begin{table*}[!t]
  \renewcommand{\arraystretch}{1.15} 
  \centering
  \begin{small}
    \begin{tabular}{cccccccccc}
      \hline
      \hline
   {\textsf{Proj}\,$\longrightarrow$\,\textsf{Frag}} & {\GALPROP} & {\WNEW} & {\YIELDX} & {$\sigma^{\rm H}_{\rm{P}\rightarrow\rm{F}}\pm \delta\sigma^{\rm H}_{\rm{P}\rightarrow\rm{F}}$} & Data sets & $F_{\alpha/p}$ &  {$\sigma^{\rm He}_{\rm{P}\rightarrow\rm{F}}\pm \delta\sigma^{\rm He}_{\rm{P}\rightarrow\rm{F}}$} & $\xi_{\rm{P}\rightarrow\rm{F}}$ & $\eta_{\rm{P}\rightarrow\rm{F}}$ \\
   \hline
 $^{16}$\Oxy{} $\longrightarrow$ $^{11}$\B{}  & 27.34 mb & 14.57 mb & 14.41 mb & (25.66 $\pm$ 1.06) mb &  {\tiny$\triangle$}$\bullet$$\Box\ast$$\blacktriangledown$  & 1.34 & (34.37 $\pm$ 1.43) mb & 0.94\,$\pm$\,0.04 & 1.01\,$\pm$\,0.02 \\ 
 $^{16}$\Oxy{}  $\longrightarrow$ $^{10}$\B{}  & 11.00 mb & 9.52 mb & 8.81 mb & (11.92 $\pm$ 0.52) mb & {\tiny$\triangle$}$\bullet$$\Box\ast$$\blacktriangledown$   & 1.34 & (15.97 $\pm$ 0.70) mb &1.08\,$\pm$\,0.04 & 0.98\,$\pm$\,0.03 \\ 
 $^{15}$\N{}  $\longrightarrow$ $^{11}$\B{}  & 26.12 mb & 26.12 mb & 21.71 mb & (30.63 $\pm$ 2.48) mb & $\bullet$$\circ$  & 1.31 & (40.04 $\pm$ 3.24) mb & 1.17\,$\pm$\,0.08 & 0.67\,$\pm$\,0.51 \\
 $^{15}$\N{} $\longrightarrow$ $^{10}$\B{}  & 9.56 mb & 8.81 mb & 7.63 mb & (9.69 $\pm$ 0.77) mb &  $\bullet$$\circ$  & 1.31 & (12.66 $\pm$ 1.01) mb & 1.01\,$\pm$\,0.07 & 1.54\,$\pm$\,0.05 \\
 $^{14}$\N{}  $\longrightarrow$ $^{11}$\B{}  & 29.22 mb & 29.98 mb & 26.66 mb & (29.80 $\pm$ 1.08) mb &  $\bullet$$\Box$$\ast$$\star$  & 1.21 & (35.94 $\pm$ 1.30) mb & 1.02\,$\pm$\,0.03 & 0.95\,$\pm$\,0.02 \\ 
 $^{14}$\N{}  $\longrightarrow$ $^{10}$\B{}  & 10.44 mb & 10.64 mb & 9.18 mb & (10.15 $\pm$ 0.84) mb & $\bullet$$\Box$$\ast$$\star$  & 1.21 & (12.24 $\pm$ 1.01) mb &0.97\,$\pm$\,0.07 & 0.99\,$\pm$\,0.05 \\
 $^{12}$\C{} $\longrightarrow$ $^{11}$\B{}  & 56.88 mb & 54.86 mb & 52.83 mb & (54.73 $\pm$ 2.57) mb & $\bullet$$\Box${\tiny$\blacksquare$}$\ast$$\blacktriangledown$$\star${\tiny $\lozenge$}  & 1.29 & (70.50 $\pm$ 3.32) mb & 0.96\,$\pm$\,0.04 & 0.90$\pm$\,0.01 \\
 $^{12}$\C{} $\longrightarrow$ $^{10}$\B{}  & 12.30 mb & 16.21 mb & 11.59 mb & (12.05 $\pm$ 0.58) mb & $\bullet$$\Box${\tiny$\blacksquare$}$\ast$$\blacktriangledown$$\star${\tiny $\lozenge$}  & 1.29 & (15.52 $\pm$ 0.74) mb & 0.98\,$\pm$\,0.04 & 1.00\,$\pm$\,0.02 \\
   \hline
$^{16}$\Oxy{} $\longrightarrow$ $^{10}$\Be{}  & 2.14 mb & 1.34 mb & 2.07 mb & (1.90 $\pm$ 0.13) mb & {\tiny$\triangle$}$\blacktriangledown$ & 1.47 & (2.79 $\pm$ 0.19) mb &    0.88\,$\pm$0.05 & 0.90\,$\pm$\,0.01 \\
$^{16}$\Oxy{} $\longrightarrow$ $^{9}$\Be{}  & 3.48 mb & 3.35 mb & 3.51 mb & (3.40 $\pm$ 0.22) mb & {\tiny$\triangle$}$\ast$$\blacktriangledown$ & 1.47 & (4.99 $\pm$ 0.32) mb & 0.97\,$\pm$\,0.06 & 0.98\,$\pm$\,0.03 \\
$^{16}$\Oxy{} $\longrightarrow$ $^{7}$\Be{}  & 10.00 mb & 8.75 mb & 8.92 mb & (8.97 $\pm$ 0.29) mb & {\tiny$\triangle$}$\blacktriangledown$ & 1.47 & (13.16 $\pm$ 0.42) mb & 0.89\,$\pm$\,0.03& 0.98\,$\pm$\,0.02 \\
$^{14}$\N{} $\longrightarrow$ $^{10}$\Be{}  & 1.75 mb & 1.06 mb & 1.81 mb & (1.73 $\pm$ 0.21) mb & {\tiny$\triangle$}{$\Box$} & 1.43 & (2.47 $\pm$ 0.30) mb & 0.99\,$\pm$\,0.09 & 0.95\,$\pm$\,0.07 \\
$^{14}$\N{} $\longrightarrow$ $^{7}$\Be{}  & 10.10 mb & 7.46 mb & 8.47 mb & (7.90 $\pm$ 0.47) mb & {\tiny$\triangle$}$\blacktriangle${$\Box$} & 1.43 & (11.29 $\pm$ 0.67) mb & 0.78\,$\pm$\,0.04 & 1.05\,$\pm$\,0.08 \\
$^{12}$\C{} $\longrightarrow$ $^{10}$\Be{}  & 3.94 mb & 2.05 mb & 3.41 mb & (3.61 $\pm$ 0.27) mb & {\tiny$\triangle$}$\Box${\tiny$\blacksquare$}$\blacktriangledown${\tiny $\lozenge$} & 1.41 & (5.07 $\pm$ 0.38) mb & 0.91\,$\pm$\,0.06 & 0.92\,$\pm$\,0.01 \\
$^{12}$\C{} $\longrightarrow$ $^{9}$\Be{}  & 6.76 mb & 5.31 mb & 4.98 mb & (6.63 $\pm$ 0.29) mb &  {\tiny$\triangle$}$\Box${\tiny$\blacksquare$}$\ast$$\blacktriangledown${\tiny $\lozenge$}$\blacktriangle$ & 1.41 & (9.32 $\pm$ 0.41) mb & 0.98\,$\pm$\,0.04 & 1.00\,$\pm$\,0.02 \\
$^{12}$\C{} $\longrightarrow$ $^{7}$\Be{}  & 9.58 mb & 10.32 mb & 10.76 mb & (8.88 $\pm$ 0.30) mb & {\tiny$\triangle$}$\bullet$$\Box${\tiny$\blacksquare$}$\blacktriangledown$$\star${\tiny $\lozenge$} & 1.41 & (12.48 $\pm$ 0.42) mb & 0.93\,$\pm$\,0.03 & 1.09\,$\pm$\,0.19 \\
\hline
$^{11}$\B{} $\longrightarrow$ $^{10}$\B{}  & 38.91 mb & 42.58 mb & 38.91 mb & (37.83 $\pm$ 9.25) mb & $\bullet$$\circ$ & 1.29 & (48.63 $\pm$ 11.89) mb & 0.97\,$\pm$\,0.21 & 1.10\,$\pm$\,0.14 \\
$^{11}$\B{} $\longrightarrow$ $^{10}$\Be{}  & 12.95 mb & 5.90 mb & 4.56 mb & (7.39 $\pm$ 0.90) mb & $\bullet$$\Box$$*$$\times$   & 1.40 & (10.36 $\pm$ 1.26) mb & 0.57\,$\pm$\,0.06 & 0.90\,$\pm$\,0.17 \\
$^{11}$\B{} $\longrightarrow$ $^{9}$\Be{}  & 10.00 mb & 15.27 mb & 8.01 mb & (7.22 $\pm$ 1.08) mb & {\tiny$\triangle$}$\bullet$ & 1.40 & (10.13 $\pm$ 1.51) mb & 0.72\,$\pm$\,0.09 & 0.91\,$\pm$\,0.12 \\
$^{11}$\B{} $\longrightarrow$ $^{7}$\Be{}  & 4.48 mb & 4.48 mb & 3.63 mb & (4.68 $\pm$ 0.49) mb &  $\bullet$$\blacktriangle$$*$  & 1.40 & (6.57 $\pm$ 0.69) mb & 1.05\,$\pm$\,0.10 & 0.90\,$\pm$\,0.11 \\
\hline
    \end{tabular}
    \caption{Production \XSs{} at $E=$\,10\,GeV/n from the existing formulae and results from the re-fitting procedure.}\label{Tab::CrossSectionErrors}
  \end{small}
\end{table*}
%
The parameters $\xi$ and $\eta$ represent normalization and energy scale, respectively, and are determined from the available data for each \PF{} channel.
For some channels, measurements are available only in narrow energy ranges so that, for these reactions, the parameter $\eta$ is poorly determined.
Nonetheless, in all channels, the procedure returns new \XSs{} along with uncertainties corresponding to one-sigma confidence intervals.
The fitting results are summarized in Table\,\ref{Tab::CrossSectionErrors}, where the renormalized \XSs{} 
are compared with the predictions of other algorithms at $E=$\,10\,GeV/n.
The best fit parameters $\xi$ and $\eta$ are also listed.
The symbols corresponding to the various data sets are encoded in Sec.\ref{Sec::Observations}. 
It can be noted that the renormalized \XSs{} are often close to the original $\sigma^{P\rightarrow F}_{\rm G}$ values,
because the original \GALPROP{} model is built with the help of a large set of data.
In contrast, the \WNEW{} and \YIELDX{} models show large discrepancies for some reactions
(\eg,  $^{16}$\Oxy{} $\rightarrow$ $^{11}$\B{} or $^{11}$\B{} $\rightarrow$ $^{10}$\Be{}).
In fact they are normalized to incomplete sets of data.
If one switches among the different \XS{} models, the resulting discrepancy amounts to about 20\,\% \citep{Maurin2010}. 
But this discrepancy is not fully representative of the systematic errors in \XS, because the available data give tighter constraints.
For instance, if the fit procedure is repeated using \WNEW{} or \YIELDX{} as the base, the results agree well
at the percent level at energies above a few 100\,MeV/n. Larger discrepancies persist in the lowest energy region ($E\lesssim$\,100-200\,MeV/n),
because at these energies many \PF{} reactions are resonant shaped, but this feature is accounted for only by the \GALPROP{} \XSs.

A subdominant contribution comes from CR spallation off helium target, which constitutes a 10\% fraction of interstellar gas.
To account for this component, all the production and destruction \XSs{} off \He{} target are obtained using  the scaling
factor $F_{\alpha/p}$ \citep{Ferrando1988}, which ranges from $\sim$\,1.2 to 1.5.
This factor is listed in Table\,\ref{Tab::CrossSectionErrors}, along with the \XS{} values for CR collisions off the \He{} target.
Due to the lack of data, uncertainties \XSs{} reactions off helium cannot be directly estimated via a measurement-validated procedure.
In this work, we conservatively assume a full correlation with the uncertainties in hydrogen production.
Improving measurements on nucleus-helium collisions is clearly helpful for reducing nuclear uncertainties, but not critical.
Reactions off \He-target give a second order contribution to the total uncertainty on the secondary production terms. 
Finally, CR propagation models require estimates of destruction processes,
for which  we have employed dedicated parametrizations \citep{Allison2016}.
These reactions are known with better precision in comparison to partial fragmentation reactions.
Furthermore, it can be estimated that only a small fraction ($\lesssim$10\,\%) of
light CR nuclei ($Z\lesssim$8) is involved in destructive processes, because the
diffusion timescale of these particles is always dominant over spallation \citep{Tomassetti2015Upturn}.
This translates into negligible uncertainties for the \Be-\B{} equilibrium fluxes.

\begin{figure*}[!t]
\begin{center}
  \includegraphics[width=0.95\textwidth]{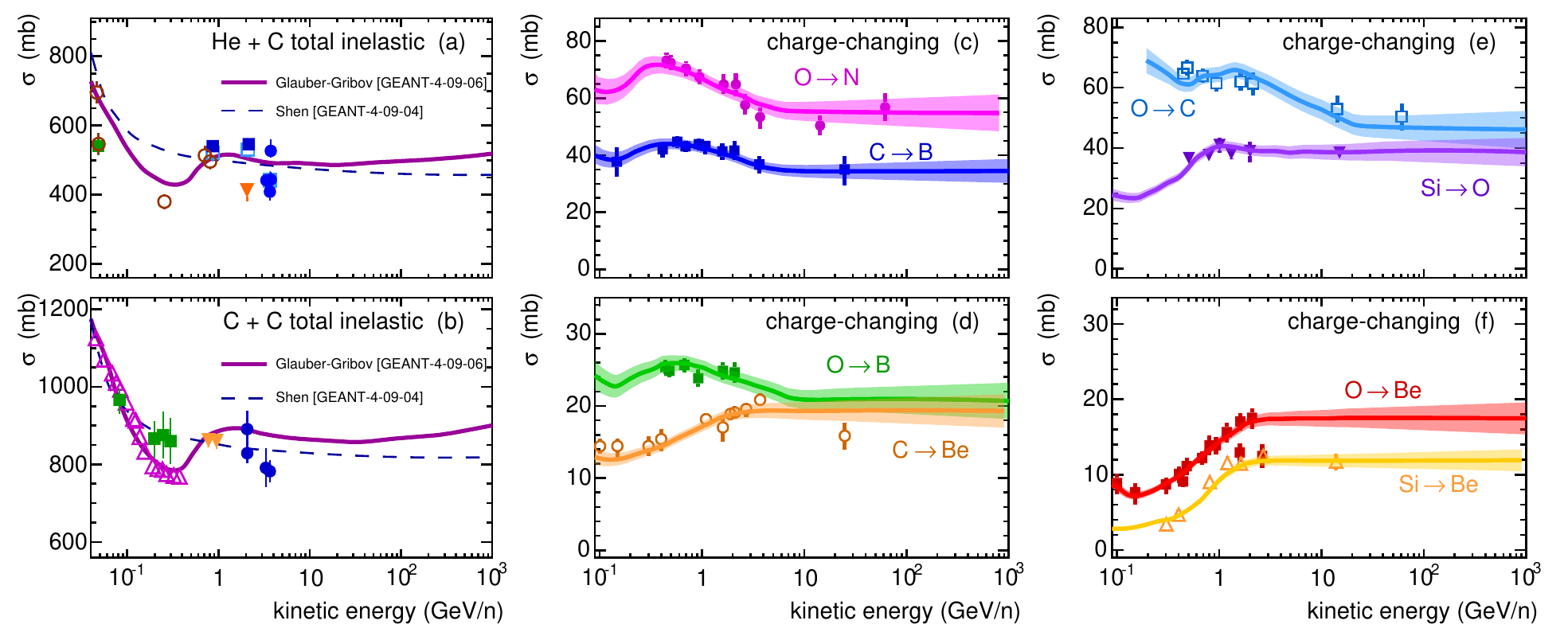}
\caption{ 
  From (a) to (b): total cross-sections for nucleus-nucleus inelastic collisions of \C{} and \He{} off carbon target \citep{Allison2016}.
  From (c) to (f): fragmentation cross-sections for charge-changing reactions off hydrogen target \citep{Webber2003}.
}\label{Fig::ccChargeChangingXS}
\end{center}
\end{figure*}

An important issue that has to be examined
is the influence of systematic energy biases in the \XS{} reactions.
In the existing parametrizations, all \XSs{} are assumed to be asymptotically energy independent above the energy of a few GeV/n energies.
Models of CR propagation rely heavily on these extrapolations. 
However, the possible --unaccounted-- presence of energy-dependent biases in these formulae may cause 
corresponding biases on the calculated \BC{} ratio which, in turn, would lead to a misdetermination of the CR transport parameters. 
For instance, energy-dependent bias in the boron production \XS{} would directly affect the \BC-driven determination of the scaling parameter $\delta$.
Examples of this effect were illustrated in \citet{Maurin2010}, and it was noticed that the application of a 
factor $E^{x}$ to the \WNEW{} parametrization is fairly consistent with the data for $|x|\lesssim$\,0.05.
A slow increase of nucleus-nucleus \XSs{} at high energy is also expected from recent developments 
based on the Glauber-Gribov model, and experimentally measured in proton-nucleus reactions.
An example is shown in Fig.\,\ref{Fig::ccChargeChangingXS} for \He+\C{} and \C+\C{} reactions,
where calculations based on the Glauber-Gribov model are compared with older calculations used 
in the \texttt{GEANT-4} hadronic generator \citep{Allison2016}.
Energy-dependent biases are not accounted for in the fitting method of Eq.\,\ref{Eq::XSRefit},
so that it returns energy independent errors.
On the other hand, the presence of high-energy biases cannot be tested for the isotopically resolved channels of 
Table,\,\ref{Tab::CrossSectionErrors}, due to the lack of multi-GeV data.
To tackle the issue, we make use of data on \emph{charge-changing} reactions that are available to nearly 100\,GeV/n of kinetic energy.
Owing the scarcity of these data, we accumulate all charge-changing measurements involving light elements from $Z=4$ to $Z=14$.
We compute the discrepancy $\Delta\sigma/\sigma$ between \XS{} data and best fit calculations as function of kinetic energy. 
Then, we adopt a $\chi^{2}$ criterion to determine the size of the allowed bias on statistical basis. 
From the relation $\delta\sigma/\sigma=\,B\log_{10}(E/E_{0})$ with $E_{0}=$\,0.1\,GeV/n,
we obtain $B=0.002\pm\,0.033$.
Thus, \XS{} data are essentially consistent with a constant behavior ($B=0$), bounded by a one-sigma 
error of $\delta{B}=0.033$ which sets the level of energy-dependent systematic uncertainty.
Charge-changing \XSs{} are shown in Fig.\,\ref{Fig::ccChargeChangingXS} (c to f) for various reactions involving
\Be{} to \Si{} nuclei, with the estimated uncertainty band.
Each charge-changing reaction reflects the combination of several isotopic channels, where 
each channel carries its own errors on normalization and energy scale.
Although these measurements can be partially correlated among channels,
(due to, \eg, common sources of systematic errors from the experimental setup),
this information is not available to us.
In contrast, energy-dependent systematics are assumed to be ``universal'' for all reactions
because, with the existing data, we were unable to perform a channel-by-channel determination.
Nonetheless, we expect this contribution to be highly correlated among the various channels, 
featuring common dynamical aspects of nucleus-nucleus collisions \citep{Allison2016}.

\section{Results and discussions} 
\label{Sec::Results}              
%
Calculations on CR modulation (Sec.\,\ref{Sec::SolarModulationCalculations}) and nuclear fragmentation
(Sec.\,\ref{Sec::CrossSectionCalculations}) are now utilized to provide final uncertainty estimates for the \BC{} ratio.
From the procedure outlined in Sec.\,\ref{Sec::SolarModulationUncertainties}, time-dependent uncertainties from
solar modulation can be obtained for the CR fluxes of any nuclear species.
boron and carbon fluxes are calculated by performing a time average over the \AMS{} observation period $\Delta{T}$, from 
May 2011 to May 2016, and using time bins of $\delta t_{i}\approx$\,1\,month:
\begin{equation}\label{Eq::NonLinearity}
J_{j}(E) = \frac{1}{\Delta T}\sum_{t_{i}\in \Delta{T}} \delta t_{i} \mathcal{\hat{G}}_{t_{i}}\left[ J_{j}^{\rm IS}\right] 
\end{equation}
A similar procedure was used to obtain a time averaged and energy-dependent error band $\delta_{\rm B/C}(E)$ for 
the modulated \BC{} ratio near Earth.
To assess the influence of the estimated \XS{} errors in the near-Earth fluxes and in the \BC{} ratio,
we proceed in calculating the uncertainties in the source terms of secondary isotopes:
\begin{equation}\label{Eq::ErrorPropagation}
  \left( \delta Q_{j}^{\rm sec} \right)^{2} = c \sum_{i}n_{i}\sum_{k} \beta_{k} N_{k} \left( \delta\sigma^{i}_{k\rightarrow j} \right)^{2} 
\end{equation}
Along with the source term, the estimated uncertainties have to be propagated to the near-Earth equilibrium fluxes $J_{j}$ 
after accounting for the effects of Galactic transport and Heliospheric modulation.
To first approximation, one has $\delta J_{j}/J_{j} \approx \delta Q^{\rm sec}_{j}/Q^{\rm sec}_{j}$ at any energy $E$,
although the connection between  $Q^{\rm sec}_{j}$ and $J_{j}$ is slightly blurred by ionization, adiabatic cooling, or reacceleration effects.
Thus we have performed a large number of propagation runs where the production terms of all secondary CRs are randomly varied, 
using Monte Carlo generation techniques, according to the Gaussian-like probability distribution of width ${\delta}Q_{j}$.
This provides Gaussian uncertainties for the secondary CR fluxes and in the \BC{} ratio calculated at different energies.
Typical uncertainties are found to be $\sim$\,5-8\,\% for \B{} production in the $\sim$\,10\,GeV/n energy scale.
All errors increase with energy because, as discussed in Sec.\,\ref{Sec::CrossSectionUncertainties},
we have accounted for the possible presence of a common energy-dependent bias.

\begin{figure}[!t]
\includegraphics[width=0.48\textwidth]{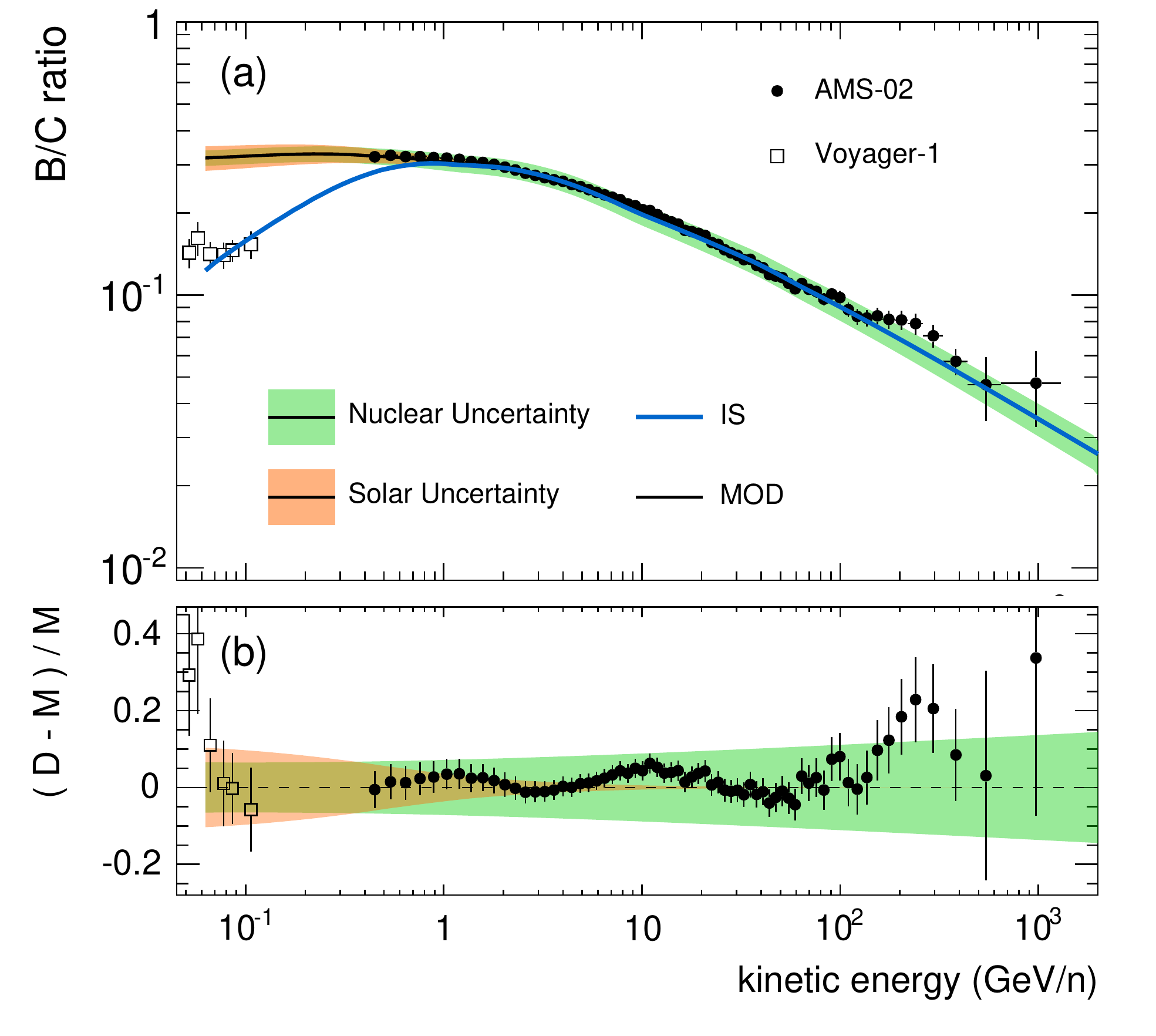}
\caption{ 
  Reference model calculations for the \BC{} ratio after accounting for (a) solar modulation, and (b) residuals between data and reference models.
  Estimates for solar and nuclear uncertainties are shown as shaded bands.
}\label{Fig::ccBCRatioResults}
\end{figure}

\begin{figure}[!t]
\includegraphics[width=0.48\textwidth]{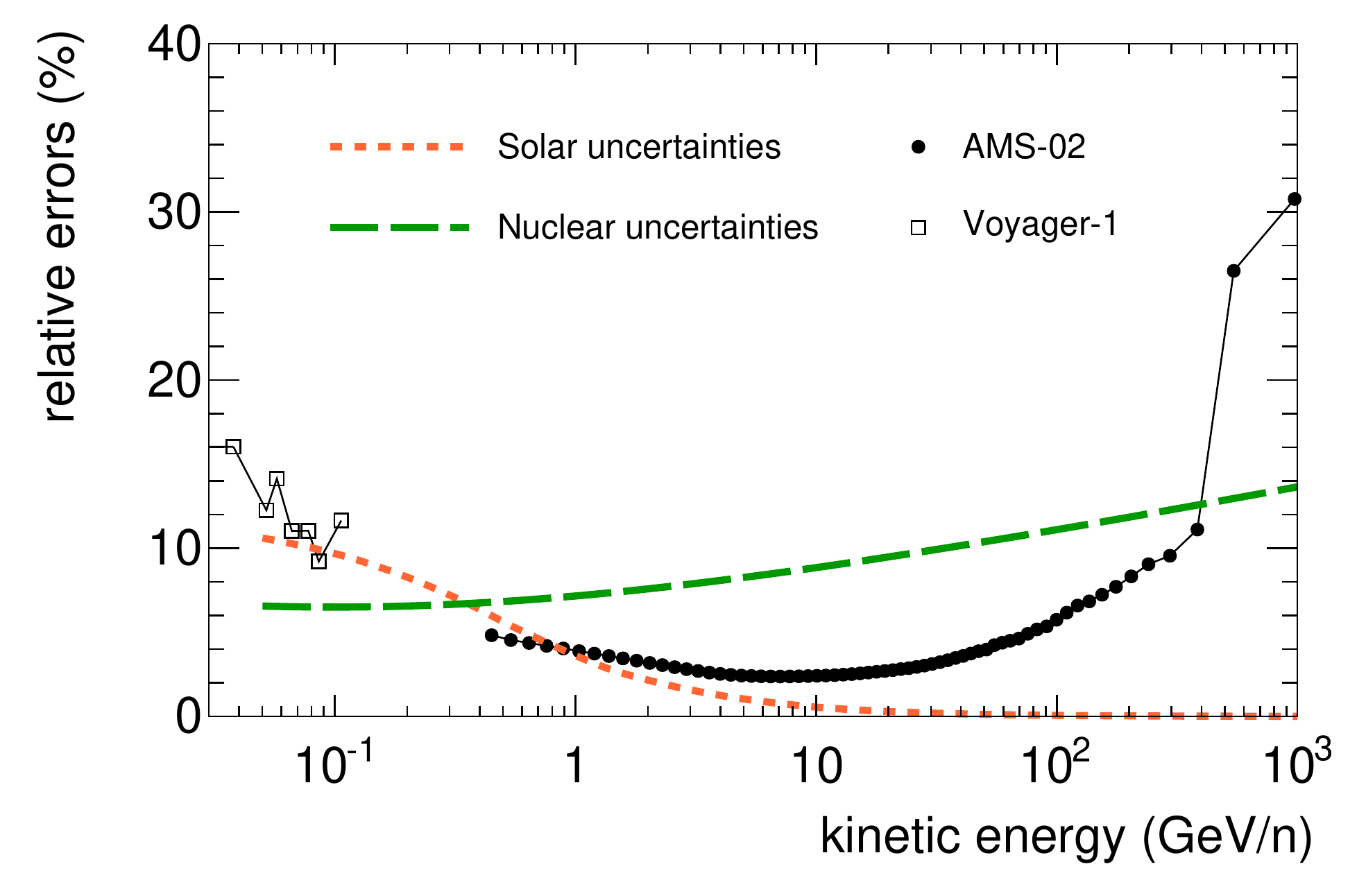}
\caption{ 
  Relative uncertainties on the \BC{} ratio arising from solar modulation and fragmentation cross-sections, 
  along with experimental uncertainties of direct \BC{} measurements from \AMS{} and Voyager-1.
}\label{Fig::ccBCRatioRelativeErrors}
\end{figure}

The solar-modulated \BC{} ratio is shown in Fig.\,\ref{Fig::ccBCRatioResults}a along with the two uncertainty bands.
In Fig.\,\ref{Fig::ccBCRatioResults}b we plot the residuals between data ($D$) and model ($M$), \ie, the quantity $(D-M)/M$.
Although the physical modeling of Galactic CR processes may be improved in several directions, we note, from this figure,
that the discrepancy between data and model lies fairly well within the systematic uncertainties.
At $E\gtrsim$\,100\,GeV/n, the \BC{} ratio has a slight tendency to harden which is not recovered by the model
and may point to unaccounted effects in CR propagation \citep{Genolini2017}. 
In the low energy region, the comparison of Fig.\,\ref{Fig::ccBCRatioResults}b
with the data from Voyager-1 is made using IS flux calculations.
With decreasing energy, below 1\,GeV/n, the ratio decreases rapidly
due to reacceleration or possible changes in diffusion \citep{Ptuskin2006}.
Our calculations break at  $60$\,MeV/n. Below these energies, and  down to $E\sim$\,8\,MeV/n, 
Voyager-1 measured a nearly energy independent \BC{} ratio that is at tension with CR propagation models \citep{Cummings2016}. 
This behavior can be further investigated with complementary data, such as charge and isotopic composition measurements
on $^{3,4}$\He, $^{1,2}$\Hyd{} or \Li-\Be, that are currently being measured by \AMS. 
It worth noticing, in fact, that the Voyager-1 probe is better optimized for the energy determination of low charge nuclei.
%
%
The relative errors on the \BC{} ratio associated with both contributions
are plotted in Fig.\,\ref{Fig::ccBCRatioRelativeErrors}.
Uncertainties from solar modulation (short-dashed lines) are estimated at the level of $\sim$\,10\,\% in the $\sim$\,100\,MeV/n energies,
then rapidly decreasing with increasing kinetic energy, below 5\,\% at $E\gtrsim$\,1\,GeV/n, and below 1\,\% at $E\gtrsim$\,10\,GeV/n.
This trend reflects the characteristic of particle transport in the heliosphere, the effects of which become weaker and weaker 
with the increasing energies of CR particles.
Uncertainties arising from nuclear fragmentation (long-dashed lines) are found to follow an opposite behavior.
These errors lie at the level of $\sim$\,6-8\,\% in the low energy region, \ie, where \XS{} data are available,
and then following a progressive increase with energy that reflects the lack of high-energy measurements.
In the figure, these curves are compared with the experimental uncertainties from the \AMS{} and Voyager-1 measurements on the \BC{} ratio.
The \AMS{} data (filled circles) can be regarded as the potential level of precision to which CR injection and transport can be understood. 
Such a potential has seen a dramatic improvement in precision and energy range, thanks to \AMS,  
turning from $\sim$\,15\,\% (in pre-\AMS{} data) to the level $\sim$\,3\,\% at $E\approx$\,50\,GeV/n of kinetic energy per nucleon.
As seen in the figure, it is reassuring to notice that solar physics uncertainties become rapidly subdominant at $E\gtrsim$\,1\,GeV/n.
Also, in the energy window $E\sim$\,0.1-1\,GeV/n, the experimental errors on the \BC{} provided by other experiments \citep{George2009}
are larger than those from solar modulation.
The experimental errors from Voyager-1 (open squares) are shown in the figure for reference, as the uncertainty line is calculated for \AMS.
Furthermore, Voyager-1 data represent a direct measurement of the IS ratio \BC. 
These data are extremely useful, \eg, to rule out CR propagation scenarios without modeling the effects of solar modulation.

The major limiting factor is represented by nuclear fragmentation inputs that dominate the uncertainties up to $\sim$\,500\,GeV/n of kinetic energy.
The highest energy points of the \AMS{} data are statistically dominated; thus we expect that the experimental errors will be reduced after longer exposure times.
This region will also be covered, very soon, from the expected release of TeV/n nuclear data by the space experiments CALET, DAMPE or ISS-CREAM \citep{Serpico2015}.

A further point of future discussion (not inspected in this work) is the validity of the straight-ahead approximation, 
according to which a narrow distribution of the fragments energies is assumed. In previous studies on the \Li-\Be-\B{} production,
the change introduced by this approximation was found to be small in comparison with the precision of the data \citep{Kneller2003}.
Within the accuracy of the new data, however, the legitimacy of this approximation may deserve a reevaluation.

Clearly, both sources of model uncertainties considered in this work affect the evaluation of the astrophysical background for dark matter searches, \ie,
the secondary fluxes of \eplus, \pbar, or \dbar{} particles that have to be calculated with \BC-driven models of CR propagation.
Along with \BC-driven uncertainties, however, astrophysical background calculations are directly affected 
by uncertainties on production and transport of antiparticles.
In particular, modeling solar modulation of antinuclei demands to account for charge-sign dependent effects (and uncertainties).
Thus, a numerical model of solar modulation including for drift motion, such as the one utilized here, is essential to get reliable estimates.
Similarly, nuclear uncertainties on the production of antiparticles are  an important factor, for estimating the background level,
that is getting increasing attention \citep{Feng2016,Kappl2015}.

\section{Conclusions}      
\label{Sec::Conclusions}   
%
The investigation of energy spectra and composition of CRs addresses fundamental science questions: 
\emph{origin of CRs: source properties, nature of injection and acceleration processes; CR diffusion properties and interstellar 
turbulence;  spatial extent and properties of the CR confinement region; presence of dark matter signatures in the CR spectrum}.
In spite of much effort in the development of CR propagation models, our knowledge of the underlying physics processes has been for long time 
plagued by large uncertainties and unresolved tensions between inferred parameters and theoretical expectations.
The need for high-energy data on secondary-to-primary ratios has often been invoked in literature \citep{Grenier2015}. 
In this respect, the past years can be considered a golden age for CR measurements. 
In comparison with previous measurements, the new \AMS{} data on the \BC{} ratio
provide a dramatic improvement corresponding to a factor $\sim$\,5  in the experimental accuracy.
To exploit these new data, however,  it is very important to precisely model the effect of solar modulation and to assess the corresponding uncertainties.
Fortunately, a large wealth of useful data on CR modulation has recently become available \citep{Bindi2017}.
New releases of local IS data and monthly resolved data of the CR flux are of invaluable help for modeling solar modulation.
Using these observations in combination with the NM counting rates, we have developed a data-driven 
approach using a numerical model of solar modulation in order to predict the near-Earth \BC{} ratio  and to assess its uncertainties.
The input parameter of our model consist in a time-series of measurements on the HCS tilt angle and NM rates
(where the latter have been converted into a more practical time-series of modulation potential $\phi$), hence our model is highly predictive. 
With the $\phi$ time-series, one may also employ the simple FF approximation for describing the data, but as discussed, the FF 
breaks down at low energies or in $A<0$ condition. Direct FF fits to CR data provides a fairly good description, but this approach 
introduces degeneracies for the CR astrophysics investigation and falls short of predicting antiparticle fluxes.
The precision of new CR data demands a reliable modeling of solar modulation.
It is also important to stress that the CR data reported by new-generation experiments are collected over ultra-long exposures 
(\eg, 5 years for the \AMS{} \BC{} ratio) during which the solar activity evolves appreciably.
Rather than using a static picture for the heliosphere with fixed parameters describing mean diffusion properties or tilt angle,
we have adopted a quasi-steady state approach where a time-series of input parameters is converted into 
a time-series of modulated CR fluxes, to be subsequently averaged over the desired observation period.
It is therefore essential, for a reliable modeling of the modulation effect, having time-resolved data over the period of interest.
In this respect, the present work would strongly benefit from monthly resolved measurements of low energy fluxes
over the period of the \AMS{} observations. 
The \AMS{} experiment is now probing the descending phase of the solar activity cycle toward the next minimum. 
Measurements in solar minimum conditions can be very precious for CR propagation studies, 
because the modulation effect during this period can be more reliable modeled. 
Another important point is that solar-modulation uncertainties are much larger for individual \B-\C{} flux calculations,
while the effect is partially suppressed when analyzing the ratios of these fluxes.
More in general, uncertainties from solar modulation are minimized with the use of secondary-to-primary 
ratios between species of similar charge or mass, \eg, \d/\p, $^{3}$\He/$^{4}$\He, \BC, and \F/\Ne.

The other important source of uncertainty is that related to nuclear fragmentation processes of CRs in the ISM.
Models of CR propagation rely heavily on high-energy extrapolations of parametric formulae for total and partial reactions.
Cross-section formulae are based on low energy nuclear data (up to a few GeV/nucleon) mostly collected in the 
90's. These data are essentially untested in the energy range relevant for the current CR physics investigation. 
In this work, we have performed a data-driven reevaluation of isotopic \XS{} data for several \PF{} channels involving the production of \B-\Be{} nuclei. 
Along with normalization and energy scale, we have accounted for uncertainties from energy-dependent biases in \XS{} formulae.
For this task we used data on charge-changing reactions.
As we have shown, nuclear-physics uncertainties are found to be a major limiting factor for the interpretation of the new \BC{} data from \AMS.
Hence, in the analysis of CR data, these uncertainties cannot be ignored any longer.
In line with recent studies, all calling for more precise nuclear data for CRs 
\citep{Gondolo2014,Maurin2010,Moskalenko2013,Tomassetti2015XS,Tomassetti2012Isotopes},
we therefore stress once more the urgent need for establishing a program 
of cross-section measurements at the $\mathcal{O}$(100\,GeV) energy scale.
To conclude, we would like to quote \citet{Chen1997} from the Transport Collaboration:
\emph{
``With the shutdown of the LBL Bevalac and the pending closure of the Saclay Saturne accelerators, opportunities for obtaining 
cross-section measurements relevant to the interpretation of CR data are rapidly dwindling worldwide. 
Thus, future experiments will rely heavily upon cross-section predictions, and it is important to update our formulae using data (...) 
to ensure that the solutions to some astrophysical problems are not dominated by cross-section inaccuracies rather than by CR measurements''.}
Nowadays, the solutions to topical problems in CR physics seems actually \emph{dominated by \XS{} inaccuracy rather than by CR measurements}.
\\[0.15cm]
\footnotesize{%
  I thank David Maurin for sharing the code \USINE{} of CR propagation calculations. 
  This work has also benefit from the public codes
  \GALPROP{} (for fragmentation \XSs) and \SOLARPROP{} (for solar modulation).
  CR data are retrieved from online databases SSDC-DB at the {\it Italian Space Agency} and CRDB at LPSC-Grenoble \citep{Maurin2014}.
  NM data are taken from NMDB \citep{Steigies2015}. 
The author acknowledges support from MAtISSE.
This project has received funding from the European Union's Horizon 2020 research and innovation programme under the Marie Sklodowska-Curie grant agreement No 707543.
}



\end{document}